\documentclass[a4paper,11pt]{article}

\usepackage[utf8]{inputenc}
\usepackage[english]{babel}
\usepackage[myheadings]{fullpage}
\usepackage[T1]{fontenc}
\usepackage{subfig}

\usepackage{graphicx}
\usepackage{sectsty}
\usepackage{url}
\usepackage[dvipsnames, table, xcdraw]{xcolor}
\usepackage{amsmath}
\usepackage{amssymb}
\usepackage{calrsfs}
\usepackage{bbold}
\usepackage{dsfont}
\usepackage{esint}
\usepackage{indentfirst}
\usepackage{xfrac}
\usepackage{bigints}
\usepackage{slashed}
\usepackage{breqn}
\usepackage{autobreak}
\usepackage{setspace}
\usepackage[colorlinks]{hyperref}
\usepackage{microtype}
\usepackage{quoting}
\usepackage{float}
\usepackage{array}
\usepackage[export]{adjustbox}
\usepackage[section]{placeins}
\usepackage[capitalise]{cleveref}
\usepackage{enumitem}
\usepackage{scalefnt}
\usepackage{booktabs}
\usepackage{caption}
\usepackage{makecell}
\usepackage{authblk}
\usepackage{cases}
\usepackage{multirow}
\usepackage[a]{esvect}

\usepackage[normalem]{ulem}

\usepackage{cite}

\usepackage{xspace}

\usepackage{colortbl}
\definecolor{myblue}{rgb}{0.8,0.85,1}
\definecolor{light-gray}{gray}{0.95}

\hypersetup{colorlinks=true,
citecolor=Magenta,
anchorcolor=CornflowerBlue,
linkcolor=NavyBlue,
urlcolor=Mahogany,
linkbordercolor={1 1 0}}

\usepackage{orcidlink}

\DeclareMathAlphabet{\mathcal}{OMS}{cmsy}{m}{n}

\makeatletter
\newcommand{\thickhline}{\noalign {\ifnum 0=`}\fi \hrule height 1pt
\futurelet \reserved@a \@xhline}
\newcolumntype{"}{@{\hskip\tabcolsep\vrule width 1pt\hskip\tabcolsep}}
\makeatother

\allowdisplaybreaks

\def\beq {\begin{equation}}
\def\eeq {\end{equation}}
\def\bea {\begin{eqnarray}}
\def\eea {\end{eqnarray}}
\def\ni {\noindent}
\def\nn {\nonumber}
\def\MSbar{\overline{\rm MS}}
\newcommand{\sand}[3]{\langle {#1}|{#2}|{#3}\rangle}

\def\amuHVP{a_\mu^{\rm HVP}}

\AtBeginDocument{}

\title{\LARGE {\bf \boldmath \sffamily Perturbative QCD below charm threshold: theory and tensions with $e^+e^-$ data}}

\author{Diogo Boito\,\orcidlink{0000-0002-4426-7984}}
\author{Marcelle Caram\,\orcidlink{0000-0002-8847-0159}}

\affil{\it Instituto de Física de São Carlos, Universidade de São Paulo, IFSC –
USP, 13566-590, São Carlos, SP, Brasil.\vspace{0.2cm}}

\date{}

\begin{document}
\begin{flushright}
{\small \today}
\end{flushright}

\begingroup
\let\newpage\relax
\maketitle
\endgroup
\date{}

\begin{abstract}
\noindent In this paper we carefully assess the theory prediction for $R(s)$ below charm threshold, $R_{uds}$, and address tensions with the existing data, notably with the 2021 BES-III results. We analyze the uncertainty of the perturbative QCD description in the light of renormalons
making use of the large-$\beta_0$ limit and the
renormalon-free gluon-condensate scheme. We provide a reliable estimate of the
duality violation contributions; we show they are sizable up to $2.5$~GeV and improve the agreement between theory and data, but are negligible for higher energies. We then combine the available experimental data for $R_{uds}$ and find the data sets to be mutually compatible.
Finally, we compare theory and data, both locally and in their contributions to the anomalous magnetic moment of the muon.
Theory is compatible with the combined data but discrepancies with the
BES-III data reach more than~3$\sigma$.
\end{abstract}

\thispagestyle{empty}

\newpage

\setcounter{page}{1}

\vspace{1cm}

\section{Introduction}

The inclusive hadronic $R(s)$ ratio
\begin{equation}
R(s) =\frac{3s}{4\pi\alpha_{\rm EM}^2}\sigma^{(0)}(e^+e^-\to {\rm hadrons}(+\gamma)),
\end{equation}
where $\alpha_{\rm EM}$ is the electromagnetic (EM) fine-structure constant and $\sigma^{(0)}$ is a bare (excluding vacuum polarization effects) photon-inclusive cross section,
has played a fundamental role in the development of QCD and is still crucial for several phenomenological applications. Experimental results for $R(s)$ are the basis for the data-driven determination of the hadronic vacuum polarization (HVP) contribution to the anomalous magnetic moment of the muon~\cite{Aliberti:2025beg}, $a_\mu^{\rm HVP}$, and remain a valuable source of information about the charm- and bottom-quark masses~~\cite{Chetyrkin:2009fv,Chetyrkin:2010ic,Dehnadi:2011gc,Dehnadi:2015fra,Erler:2016atg}, as well as about the strong coupling~\cite{Boito:2018yvl, Boito:2019pqp,Boito:2020lyp},~$\alpha_s$.
In the present paper, we address the theory description of $R(s)$ below open-charm threshold, which we denote $R_{uds}$, and the tension that has emerged in 2021, after new inclusive measurements of $R_{uds}$, between $\sqrt{s}=2.23$~GeV and $3.67$~GeV, were made public by the BES-III collaboration~\cite{BESIII:2021wib}.
The new BES-III results are rather precise, but are systematically larger than the perturbative QCD (pQCD) prediction and than some of the previous measurements of $R_{uds}$, notably those by the KEDR collaboration~\cite{KEDR:2018hhr,Anashin:2015woa,Anashin:2016hmv}.

One of the main reasons to quantitatively assess the status of perturbative QCD for $R_{uds}$ is its aforementioned connection with the anomalous magnetic moment of the muon, $a_\mu$. The new measurements of $a_\mu$, performed by the FNAL E989 experiment at Fermilab~\cite{Muong-2:2021ojo,Muong-2:2023cdq , Muong-2:2025xyk}, which are in agreement with the previous, less precise, results from the Brookhaven National Lab BNL E821 experiment~\cite{Muong-2:2006rrc}, lead to a combined world experimental average for $a_\mu$ with an uncertainty of a mere $124$ ppb.
The status of the determination of this quantity in the Standard Model (SM) has recently been reviewed in the 2025 $g-2$ Theory Initiative (TI) White Paper (WP)~\cite{Aliberti:2025beg}, based on results from Refs.~\cite{Aoyama:2012wk,Volkov:2019phy,Volkov:2024yzc,Aoyama:2024aly,Parker:2018vye,Morel:2020dww,Fan:2022eto,Czarnecki:2002nt,Gnendiger:2013pva,Ludtke:2024ase,Hoferichter:2025yih,RBC:2018dos,Giusti:2019xct,Borsanyi:2020mff,Lehner:2020crt,Wang:2022lkq,Aubin:2022hgm,Ce:2022kxy,ExtendedTwistedMass:2022jpw,RBC:2023pvn,Kuberski:2024bcj,Boccaletti:2024guq,Spiegel:2024dec,RBC:2024fic,Djukanovic:2024cmq,ExtendedTwistedMass:2024nyi,MILC:2024ryz,Bazavov:2024eou,Keshavarzi:2019abf,DiLuzio:2024sps,Kurz:2014wya,Colangelo:2015ama,Masjuan:2017tvw,Colangelo:2017fiz,Hoferichter:2018kwz,Eichmann:2019tjk,Bijnens:2019ghy,Leutgeb:2019gbz,Cappiello:2019hwh,Masjuan:2020jsf,Bijnens:2020xnl,Bijnens:2021jqo,Danilkin:2021icn,Stamen:2022uqh,Leutgeb:2022lqw,Hoferichter:2023tgp,Hoferichter:2024fsj,Estrada:2024cfy,Deineka:2024mzt,Eichmann:2024glq,Bijnens:2024jgh,Hoferichter:2024bae,Holz:2024diw,Cappiello:2025fyf,Colangelo:2014qya,Blum:2019ugy,Chao:2021tvp,Chao:2022xzg,Blum:2023vlm,Fodor:2024jyn}. Progress has been achieved in several fronts since the previous version of the $g-2$ TI WP~\cite{Aoyama:2020ynm} (notably in the lattice-QCD evaluation of $a_\mu^{\rm HVP}$) but the main source of uncertainty in the SM result remains $a_\mu^{\rm HVP}$. Using as input the lattice-QCD results for $\amuHVP$, the SM determination of~$a_\mu$ is in agreement with the experimental result within 0.6$\sigma$. The situation for the data-driven approach to $\amuHVP$, based on data for $R(s)$, is less clear, since the CMD-3 results~\cite{CMD-3:2023rfe,CMD-3:2023alj} for the cross-sections $e^+e^- \to \pi^+\pi^-$ show a strong tension with previous measurements of the same quantity. Data-driven results that use the CMD-3 cross-sections lead to agreement with lattice-QCD~\cite{Boito:2022rkw,Boito:2022dry,Benton:2023dci,Benton:2023fcv,Benton:2024kwp} and with the experimental determination of $a_\mu$~\cite{Aliberti:2025beg}, while results based on a combination of previous measurements of the same cross-sections lead to a data-driven determination of $a_\mu$ that disagrees with experimental results and show a significant tension with lattice-QCD~\cite{Benton:2024kwp,Keshavarzi:2019abf,Davier:2019can,Aliberti:2025beg}.
Finally, data-driven results based on $\tau\to \pi^-\pi^0\nu_\tau$ measurements, that must rely on non-trivial model-dependent isospin-breaking corrections~\cite{Masjuan:2023qsp,Miranda:2020wdg,Hoferichter:2023sli,Colangelo:2022prz,Davier:2010fmf,Davier:2023fpl,Davier:2025jiq,Colangelo:2025iad}, are lower but compatible with the recent lattice-QCD determinations.

In the
data-driven approach to $\amuHVP$, in the energy region of interest to this paper --- between $1.8$~GeV and $3.7$~GeV --- some works use the available experimental
data~\cite{Keshavarzi:2018mgv,Keshavarzi:2019abf} while others turn to the use of pQCD expressions~\cite{Davier:2017zfy,Davier:2019can}. The latter approach assumes, of course, that pQCD is valid and leads to a smaller final error for this contribution to $\amuHVP$, since the pQCD prediction has smaller errors than the experimental determinations of $R(s)$.
Given the size of the
contribution of this energy region to the final data-driven assessment of $\amuHVP$ ($\sim 30\times 10^{-10}$), the resolution of this tension is not nearly as crucial as the issues in exclusive measurements of $e^+e^- \to \pi^+\pi^-$.
Nevertheless, a strong disagreement between experimental results for $R_{uds}$ and the prediction in pQCD --- presently known at five loops, or $\mathcal{O}(\alpha_s^4)$~\cite{Baikov:2008jh,Herzog:2017dtz} --- besides having phenomenological implications, is disconcerting at a conceptual level since, as we will
discuss, the uncertainties on the pQCD theory prediction are very small and a large portion of this energy region is sufficiently far from resonances for one to expect that
pQCD should already provide a good description of $R(s)$ measurements, especially for the higher-energy portion of the $R_{uds}$ domain. Furthermore, a discrepancy between theory and $R_{uds}$ measurements has implications for $\alpha_s$ and charm-quark mass determinations from relativistic sum rules, since in those studies the $R_{uds}$ background must be subtracted from the inclusive four-flavor $R(s)$ measurements above charm threshold~\cite{Chetyrkin:2009fv,Chetyrkin:2010ic,Dehnadi:2011gc,Dehnadi:2015fra,Erler:2016atg,Boito:2020lyp,Boito:2019pqp}.

Our aim in this paper is to, first, carefully assess the present knowledge of the pQCD description of $R_{uds}$, with a reliable estimate of uncertainties due to missing higher orders. Potential convergence issues in the pQCD series are quantified using what is known about renormalons~\cite{Beneke:1998ui,Beneke:2008ad,Beneke:2012vb,Boito:2020hvu}, supplemented with realistic models for higher-orders~\cite{Beneke:2008ad,Boito:2018rwt}, as well as employing the recently introduced renormalon-free (RF) gluon-condensate scheme~\cite{Benitez-Rathgeb:2022yqb,Benitez-Rathgeb:2022hfj,Benitez-Rathgeb:2021gvw} and the large-$\beta_0$ limit of QCD. After discussing the small quark-mass corrections, we employ the knowledge available about quark-hadron duality violations (DVs)~\cite{Cata:2005zj,Cata:2008ye,Boito:2017cnp} in $e^+e^-\to {\rm hadrons}$~\cite{Boito:2018yvl}, as well as in the isospin-related process $\tau \to {\rm hadrons}+\nu_\tau$~\cite{Boito:2025pwg,Boito:2020xli}, to estimate this non-perturbative contribution. We show that the DVs can be sizable for $\sqrt{s}\leq 2.5$~GeV and that this contribution improves the agreement between theory and data. We are then in a position to produce state-of-the-art theory results with a realistic and reliable uncertainty both locally, for $R_{uds}(s)$, as well as in integrated results for $\amuHVP$ in the $R_{uds}$ region. We then turn to the experimental data.
With the algorithm of Refs.~\cite{Keshavarzi:2018mgv,Keshavarzi:2019abf} (see also Ref.~\cite{Boito:2025pwg}), we combine the experimental data for $R_{uds}$ to produce a combined experimental result for $\amuHVP$ in the $R_{uds}$ region while assessing the compatibility of the different experimental results.
We also compute, in two different energy windows, the contribution to $\amuHVP$ implied by individual data sets for $R_{uds}$. Finally, we
quantify the local as well as the integrated discrepancy between the pQCD prediction with and without DVs, the different experiments, and the combined experimental results.

This paper is organized as follows. In Sec.~\ref{sec:theory} we give an overview of the theoretical framework. In Sec.~\ref{sec:ptQCD} we discuss in detail the pQCD contribution and its uncertainty. Other corrections are discussed in Sec.~\ref{sec:othercorrections}, with a special focus on the DV contribution. In Sec.~\ref{sec:R-results}, we give theory predictions for $R_{uds}$ with all corrections included. In Sec.~\ref{sec:Comparison}, we combine the available data sets and perform a quantitative comparison between theory and data. Our conclusions are given in Sec.~\ref{sec:Conclusions}. Some of the details of the implementation of the RF scheme are relegated to App.~\ref{app:RFGC}.

\section{\boldmath Theoretical framework}\label{sec:theory}

We start with a brief recollection of well-known results.
The result for $R_{uds}$ in pQCD can be obtained, with the use of the optical theorem, from the
massless Adler function. Mass corrections can then be added perturbatively, but for $R_{uds}$ they are very small, as we will discuss. We remind that, because of the quark charges involved, $R_{uds}$ does not receive contributions from singlet diagrams.
Therefore, the results can be cast in terms of the vector correlator of two (massless) non-singlet quark-field currents
\beq
\Pi^V_{\mu\nu}(q^2) =(q_\mu q_\nu -g_{\mu\nu}q^2)\Pi(q^2) =i \int d^4 x\, e^{i q\cdot x} \sand{0}{j_\mu^V(x)j^V_\nu(0)^\dagger}{0},
\eeq
with $j_\mu^V(x)=\bar u \gamma_\mu d(x)$. It is customary to work with the renormalization-group invariant Adler function, $D(q^2)$, defined as
\beq
D(q^2)=-q^2\frac{d}{dq^2} \Pi(q^2)=\frac{N_c}{12\pi^2}\left(1+\widehat D(q^2) \right),
\eeq
where we also defined the reduced Adler function, $\widehat D(q^2)$, which starts at $\mathcal{O}(\alpha_s)$.

The result for $\Pi(q^2)$, or for the associated Adler function, can be organized in terms of the operator product expansion (OPE). The leading contribution, stemming from the identity operator, with dimension zero, is the purely perturbative contribution.
The general perturbative expansion of $\widehat D(s)$ in QCD is given by
\beq
\widehat D_{\rm pert}(q^2) = \sum\limits_{n = 1}^{\infty}a_s^n(\mu^2)
\sum\limits_{k = 1}^{n+1 }{k\, c_{n,k}\left[\log\left(\frac{-q^2}{\mu^2} \right)\right]^{k-1}},\label{eq:Adler-expansion}
\eeq
where $\mu$ is a renormalization scale and $a_s(\mu^2)\equiv \alpha_s(\mu^2)/\pi$. The independent coefficients that must be calculated perturbatively are the $c_{n,1}$, while the $c_{n,k>1}$ can be obtained with the renormalization group. For $R_{uds}$, and in accordance with our definitions, only the non-singlet $c_{n,k}$ coefficients intervene.
Presently, these coefficients are known exactly up to $c_{4,1}$~\cite{Baikov:2008jh,Herzog:2017dtz} (five loops). Several estimates exist for $c_{5,1}$ and higher. Here, when estimating the fifth-order term, we will use $c_{5,1}=280\pm140$, which covers the existing estimates for $c_{5,1}$ obtained with various methods~\cite{Baikov:2008jh,Beneke:2008ad,Boito:2018rwt,Caprini:2019kwp}.

The connection with $R_{uds}$ is made with the spectral function, defined as
\beq
\rho(s) = \frac{1}{\pi}{\rm Im}\Pi(s+i0),
\eeq
with $q^2=s$.
Because the correlator $\Pi(s)$ satisfies the Schwarz reflection principle, the following integral representation for the spectral function associated with $\widehat D(s)$ can be obtained with the use of Cauchy's theorem
\beq
\widehat \rho(s)= \frac{1}{2\pi i}\int\limits_{s+i0}^{s-i0}ds^\prime \, \frac{\widehat D(s^\prime)}{s^\prime} = \frac{1}{2\pi i} \oint\limits_{|x|=1} \frac{dx}{x} \widehat D(sx),
\eeq
where the integral on the left-hand side can be written as the difference of integrals over $s$ just above and below the cut by a change of variables. In the representation of the right-hand side, where $x=s^\prime/s$, which will be used from now on,
we will always consider a circular contour of integration in the complex plane with $|x|=1$. In this form, the spectral function becomes a particular case of integrated moments of the Adler function of the type
\beq
\frac{1}{2\pi i} \oint\limits_{|x|=1} \frac{dx}{x} W_m(x) \widehat D(sx)\label{eq:Adler-mom},
\eeq
with $W_\rho(x)=1$. These moments intervene in the analysis of $\alpha_s$ from hadronic $\tau$ decays and they have been extensively studied in the literature (see, e.g., Refs.~\cite{Beneke:2008ad,Beneke:2012vb,Abbas:2013usa, Caprini:2017ikn, Caprini:2018agy,Boito:2018rwt,Boito:2020hvu,Benitez-Rathgeb:2022hfj,Rodriguez-Sanchez:2025nsm}). In particular, for a long time there was a discrepancy in the results obtained from the two most widely used prescriptions for the renormalization-scale setting when using Eq.~(\ref{eq:Adler-expansion}) in the integral of Eq.~(\ref{eq:Adler-mom}) --- the question of Fixed Order Perturbation Theory (FOPT)~\cite{Beneke:2008ad} versus Contour Improved Perturbation Theory (CIPT)~\cite{LeDiberder:1992jjr,Pivovarov:1991rh}. This discrepancy, unlike what is expected for a residual renormalization-scale dependence, did not become smaller with the calculation of the $\mathcal{O}(\alpha_s^4)$ coefficient. Results from FOPT, obtained with a fixed $\mu^2 =s$ in the integrand of Eq.~(\ref{eq:Adler-mom}), are systematically different from the results from CIPT, obtained with a running scale $\mu^2=-sx$, thereby resumming the logarithms of Eq.~(\ref{eq:Adler-expansion}).
The same ambiguity appears for the perturbative spectral function but, fortunately, the problem is now very well understood~\cite{Hoang:2020mkw,Hoang:2021nlz,Benitez-Rathgeb:2022yqb,Golterman:2023oml,Gracia:2023qdy} and it is clear that the standard CIPT prescription is not consistent with the usual OPE and must either be dropped or remedied. The discrepancy arises because CIPT always retains a sensitivity to infrared (IR) renormalons, which then leads to a systematic non-perturbative difference between the two results. The CIPT series can be fixed, for practical purposes, if the leading IR renormalon is consistently removed, for example with the use of the renormalon-free (RF) gluon-condensate (GC) scheme of Refs.~\cite{Benitez-Rathgeb:2022yqb,Benitez-Rathgeb:2022hfj,Benitez-Rathgeb:2021gvw}, leading to consistent results between the two series at higher orders.

Beyond perturbation theory in the chiral limit, the $D=2$ contributions arise from the perturbative quark-mass corrections. In our case, since the masses of the light quarks $u$ and $d$ are tiny, we can safely consider only the $m_s$ corrections. Those are known perturbatively up $\alpha_s^3$~\cite{Chetyrkin:1993hi, Chetyrkin:1990kr} and are discussed in detail in Sec.~\ref{sec:othercorrections}. As we will show, these corrections are small and mass corrections with higher dimension, starting at $\mathcal{O}(m_s^4/s^2)$, can safely be neglected.

Higher-dimension non-perturbative OPE corrections to $\Pi(s)$ can be cast in the form
\beq
\Pi(s)_{D>2}^{\rm OPE} = \sum_{k=2}^\infty\frac{C_{2k}(s)}{(-s)^k},
\eeq
where the coefficients $C_{2k}(s)$ encode both the perturbative Wilson coefficients and the condensates formed from Lorentz- and gauge-invariant operators of dimension $D=2k$. These higher-dimension corrections are, however, strongly suppressed for $\rho(s)$. The reason for that is that the $s$ dependence in the $C_{2k}$ coefficients arises solely from $\alpha_s$-suppressed logarithms from higher-order corrections. In the contour integration of Eq.~(\ref{eq:Adler-mom}), a monomial $x^{k}$ (with $k=2,3...$) in the weight function $W_m(x)$ picks up the non-$\alpha_s$-suppressed contribution with $D=2k$. Since for $\rho(s)$ the weight function is $W_\rho(x)=1$, only the $\alpha_s$-suppressed terms survive the contour integration and there is a strong suppression of all OPE condensate contributions. Neglecting these small $\alpha_s$-suppressed contributions from OPE condensates, about which little is known from exact calculations,\footnote{The exception is the gluon condensate, for which the Wilson coefficient is known to order $\alpha_s^3$~\cite{Bruser:2024zyg}. In this case, it can be explicitly verified that neglecting $\alpha_s$-suppressed contributions to polynomial moments of the Adler function is fully justified~\cite{Boito:2011qt,Boito:2024gtb}. Estimates for the contribution of $\alpha_s$-suppressed terms from $D=6$ condensates also exist~\cite{Boito:2024gtb}, albeit relying on assumptions about the different condensates that contribute in this case. These estimates also support the neglect of $\alpha_s$-suppressed terms.} is standard in integrated moments of the Adler function~\cite{Boito:2024gtb,Boito:2011qt,Rodriguez-Sanchez:2025nsm}, and we will therefore neglect them henceforth.

An additional source of non-perturbative contributions is the DVs. The OPE is strictly valid in the Euclidean. For $s$ sufficiently large, however, one expects that, far away from resonances, pQCD should provide a good description of $R(s)$. In this regime, the DV contribution quantifies the residual oscillations around pQCD due to superimposing tails of higher resonances, that can still be manifest in the data. DVs cannot be obtained from first principles, but an asymptotic parametrization for their effects can be obtained from well accepted assumptions about the QCD spectrum, such as asymptotic Regge trajectories and large $N_c$ considerations~\cite{Cata:2005zj,Cata:2008ye,Boito:2017cnp}. This type of parametrization has been used in several studies of $\tau\to {\rm hadrons}+\nu_\tau$ spectral functions~\cite{Cata:2008ye,Gonzalez-Alonso:2010kpl,Boito:2011qt,Boito:2020xli,Boito:2025pwg} as well as $e^+e^-\to {\rm hadrons}$~\cite{Boito:2018yvl,Boito:2022dry,Benton:2023dci}, and these results can be used to estimate DV effects in $R_{uds}$, as discussed in Sec.~\ref{sec:othercorrections}.

For $R_{uds}$, the relevant current is the EM current which reads
\beq
j_\mu^{\rm EM} = Q_u\,\bar u \gamma_\mu u +Q_d \,\bar d \gamma_\mu d +Q_s\,\bar s \gamma_\mu s = \frac{1}{2}(\bar u \gamma_\mu u - \bar d \gamma_\mu d)+\frac{1}{6}(\bar u \gamma_\mu u + \bar d \gamma_\mu d -2 \bar s \gamma_\mu s),
\eeq
where on the right-hand side we have given the decomposition in isospin $I=1$ and $I=0$ parts, which will be useful later on. Due to the aforementioned cancellation of singlet contributions proportional to $\left(\sum_i Q_i\right)^2$, the perturbative contribution to $R_{uds}$ is directly obtained from the non-singlet Adler function.

The observable $R_{uds}$ can then be written as
\begin{align}\label{eq:RudsNLO}
R_{uds}(s) &= 12\pi^2 \rho_{\rm EM}(s) = N_c \sum_{q=u,d,s} Q_q^2\left(1+\delta^{(0)}_{\alpha_s}+\delta_{\rm EM} +\delta_{m_q^2}+\delta_{\rm DVs}\right),
\end{align}
with the $\alpha_s$ corrections encoded in $\delta^{(0)}_{\alpha_s}$ and where $\delta_{\rm EM}$ is the leading EM correction, $\delta_{m_q^2}$ represents quark-mass corrections, and $\delta_{\rm DVs}$ the contribution from DVs. We will discuss and quantify these corrections in the remainder of the paper.

Finally, the contribution to $\amuHVP$ from the energy interval $s_1 \leq s \leq s_2$ can be expressed in terms of $R(s)$ as
\beq
\amuHVP[s_1;s_2] = \left( \frac{\alpha_{\rm EM} m_\mu}{3\pi}\right)^2\int_{s_1}^{s_2} ds\, \frac{\hat K(s)}{s^2} {R(s)},
\label{eq:amuhvp_def}
\eeq
where $m_\mu$ is the muon mass and the slowly-varying kernel function $\hat K(s)$~\cite{Brodsky:1967sr, Lautrup:1968tdb} can be found in Ref.~\cite{Aliberti:2025beg}.

\section{Perturbative QCD and the expected higher-order behavior}
\label{sec:ptQCD}

We begin with a discussion of what can be considered the standard treatment of pQCD in the case of $R(s)$.
Using FOPT, which amounts to setting $\mu^2=s$ in the integrand of Eq.~(\ref{eq:Adler-mom}), the result for the spectral function is written in terms of integrals over powers of $\log(-x)$ (which can be obtained analytically),~as
\beq
\delta^{(0)}_{\alpha_s,{\rm FO}} = \sum_{n=1}^{\infty} a^n_s(s)\sum_{k=1}^{n+1}\, k\,c_{n,k}\frac{1}{2\pi i}\oint\limits _{|x|=1}\frac{dx}{x}\log^{k-1}(-x),\label{eq:delta0}
\eeq
where the strong coupling at the scale $\mu^2=s$ is obtained with the five-loop QCD $\beta$~function~\cite{Baikov:2016tgj}.
These are the dominant corrections to $R_{uds}(s)$. The explicit results for these $\alpha_s$ corrections in FOPT, for three quark flavors ($N_f=3$) and up to $\alpha_s^5$, are
\begin{align}
\delta^{(0)}_{\alpha_s}(s)&= a_s(s) + 1.6398\,a_s^2(s)-10.284a_s^3(s)-106.88a_s^4(s) \nn\\ &+(c_{5,1}-779.58)a_s^5(s)+\cdots.\label{eq:delta0-QCD-c51}
\end{align}
In the $\mathcal{O}(\alpha_s^5)$ term, we have kept explicit the contribution of $c_{5,1}$, which is not known from perturbative calculations, while the $-779.58$ includes the contributions from $c_{5,2}$, $c_{5,3}$, and $c_{5,4}$, which can be written in terms of the known coefficients $c_{n<5,1}$ as well as known $\beta$-function coefficients.
For $s=(2~{\rm GeV})^2$, with $\alpha_s(4~{\rm GeV}^2)=0.2949(61)$,\footnote{This value corresponds to the current Particle Data Group average $\alpha_s(m_Z^2)=0.1180(9)$~\cite{ParticleDataGroup:2024cfk} which we will use throughout this work.} the numerical result is
\begin{align}
\delta^{(0)}_{\alpha_s}(4~{\rm GeV^2)}&=0.09387+0.01445-0.008506-0.008298-0.0036(10)_{\rm c_{51}}+\cdots \nn \\
&=0.0879(21),
\label{eq:delta0-QCD-2GeV}
\end{align}
where we have given the numerical result for each term order by order, including our estimate of the $\alpha_s^5$ contribution and its associated error from the uncertainty in the value of $c_{5,1}$. In the total result, the uncertainty reflects that of $\alpha_s$ added quadratically with an estimate for the error associated with the truncation of the perturbative series. The latter is taken as the maximum between the error obtained from variations of $c_{5,1}$ and the one stemming from the difference between the results at $\mathcal{O}(\alpha_s^4)$ and $\mathcal{O}(\alpha_s^5)$.\footnote{This choice is conservative. We checked that renormalization scale variations lead to smaller errors than the truncation error we consider.}

It is noteworthy that the exactly known coefficients of $\delta_{\alpha_s}^{(0)}$ change sign starting at $\mathcal{O}(\alpha_s^3)$. Also, the third and fourth order terms are of similar magnitude, which is uncommon for other moments of the Adler function, with the exception of moments that have a high sensitivity to the leading IR renormalon, which entails a run-away behavior in the perturbative expansion~\cite{Beneke:2012vb}. In the case of $\widehat \rho(s)$, however, there is no such sensitivity to the leading IR renormalon, as we show next.

The perturbative QCD behavior of Adler function moments has been studied in several papers~\cite{Beneke:2008ad,Beneke:2012vb,Abbas:2013usa, Caprini:2017ikn, Caprini:2018agy,Boito:2018rwt,Boito:2020hvu,Benitez-Rathgeb:2022hfj,Benitez-Rathgeb:2022yqb}. Since the perturbative series is (at best) asymptotic, it is very common to study it in terms of its Borel transform. For $\mu^2=-q^2=Q^2$ one can rewrite the expansion of $\widehat D$, Eq.~(\ref{eq:Adler-expansion}),~as
\beq
\widehat D(Q^2)=\sum_{n=0}^\infty r_n \alpha_s^{n+1}(Q^2),
\eeq
with $r_{n}=c_{n+1,1}/\pi^{n+1}$ and define the Borel transform, which is an inverse-Laplace transform, as in Ref.~\cite{Beneke:2008ad}
\beq
B[\widehat{D}](t)\equiv\sum_{n=0}^\infty r_n \frac{t^n}{n!} .
\eeq
The Borel-sum of the series is then defined by the Laplace transform of $B[\widehat{D}](t)$
\beq
\widehat D(\alpha)=\int_0^\infty dt\, e^{-t/\alpha}B[\widehat D](t). \label{eq:Borel-sum}
\eeq
This defines the `true value' of the series, up to potential imaginary ambiguities when there are singularities (renormalons) that obstruct the integration path.

The large-$\beta_0$ limit of QCD~\cite{Beneke:1998ui} is often used as a starting point for the analysis of the series behavior at higher orders. In a nut shell, the large-$\beta_0$ limit is obtained from results taking $N_f\to \infty$ while keeping $N_f \alpha_s \sim \mathcal{O}(1)$. The quark bubble-loop corrections to the gluon propagator, which are proportional to $N_f$, must then be summed to all orders. Gluon self-interactions are introduced with the procedure known as naive non-abelianization~\cite{Broadhurst:1994se,Beneke:1994qe}, where the fermionic contribution to the leading order QCD $\beta$ function is replaced by its full QCD result, thereby effectively introducing a set of non-Abelian contributions.

In this limit, where the series is known to all orders in $\alpha_s$, and the asymptotic character of the perturbative series is manifest, the Borel transformed Adler function can be obtained exactly.
The large-$\beta_0$ result for $B[\widehat{D}](t)$ reads\cite{Broadhurst:1992si,Beneke:1992ch}
\beq
B[\widehat{D}_{L\beta_0}](u) = \frac{32}{3\pi}\left( \frac{Q^2}{\mu^2} \right)^{-u}\frac{{\rm e}^{-Cu}}{(2-u)}\sum_{k=2}^\infty \frac{(-1)^k k}{[k^2-(1-u)^2]^2} ,\label{eq:Borel-Trans-Adler-largebeta}
\eeq
where $u\equiv\beta_1 t/(2\pi)$ and $C$ is a renormalization-scheme dependent constant; in the $\MSbar$-scheme, $C=-5/3$. In this form, the renormalons, which are the singularities in $u$ associated with the factorial growth of series coefficients, can be easily studied. The IR singularities ($u>0$) always have a simple and a double pole terms, with the sole exception of the leading IR pole, at $u=2$, associated with the $D=4$ OPE contribution from the gluon condensate, which is a simple pole. IR poles also entail an imaginary ambiguity in the Borel sum, since they obstruct the integration path in Eq.~(\ref{eq:Borel-sum}) and must be circumvented, which is usually done with the principal value prescription. The UV poles ($u<0)$ all exhibit the simple plus double structure.
IR poles are associated with fixed-sign series while UV singularities generate alternating-sign coefficients. Since the pole closest to the origin is the leading UV at $u=-1$ the series, eventually, is dominated by this pole and displays sign alternation at high orders~\cite{Beneke:2008ad}.

Since in the large-$\beta_0$ limit the series is known to all orders, it is interesting to study $\delta_{\alpha_s}^{(0)}$ in this limit and confront the results with the full QCD expression. It is simple to obtain the Borel transformed $\widehat \rho_{\rm FO}$ in the large-$\beta_0$ limit using the integral representation of Eq.~(\ref{eq:Borel-sum}) in Eq.~(\ref{eq:Adler-mom}), which gives
\beq
B[\widehat{\rho}_{{\rm FO},L\beta_0}] = \frac{\sin(\pi u)}{\pi u} B[\widehat{D}_{L\beta_0}](u)
\label{eq:borelrho} .
\eeq
This result shows that the leading IR pole of $B[\widehat{D}_{L\beta_0}](u)$ is exactly canceled by a zero of $\sin(\pi u)$ while the other IR poles are all reduced from double to simple. The cancellation of the leading IR pole is crucial for a good perturbative behavior since moments of the Adler function that are sensitive to this IR pole tend to have a run-away pattern, and never stabilize~\cite{Beneke:2012vb,Boito:2020hvu,Benitez-Rathgeb:2022yqb}. Furthermore, the cancellation of all simple IR poles means that $B[\widehat \rho_{\rm FO}
](u)$ is much less singular than $B[\widehat{D}](u)$, which tends to be associated with better behaved perturbative series~\cite{Boito:2020hvu}.

Reconstructing the $\alpha_s$ expansion of $\widehat{\rho}_{\rm FO}$ in the large-$\beta_0$ limit from $B[\widehat{D}](u)$ one finds,
\begin{align}
\delta^{(0)}_{\alpha_s,L\beta_0}&= a_s(s) + 1.56\,a_s^2(s)-0.944a_s^3(s)-52.9a_s^4(s)\nn\\ &-283a_s^5(s)-2241 a_s^6(s)+\cdots
\end{align}
The leading correction is, by construction, the same as in QCD, while the higher-orders can be considered ``predictions'' in the large-$\beta_0$ limit. Comparing with the QCD result, Eq.~(\ref{eq:delta0-QCD-c51}), the large-$\beta_0$ limit reproduces very well the $a_s^2$ term and predicts correctly the signs of the other known coefficients. The $a_s^5$ term is also similar to the estimated result in Eq.~(\ref{eq:delta0-QCD-c51}). This general pattern gives us some confidence that the large-$\beta_0$ result is able to capture the essence of the QCD series. It is worth mentioning that in large-$\beta_0$ a systematic sign alternation sets in starting from the $\mathcal{O}(a_s^7)$ term. Therefore, at intermediate orders, the flip in sign is still due to a competition of IR and UV renormalons and does not yet reflect the asymptotic dominance of the leading UV renormalon, which takes over from $\mathcal{O}(a_s^7)$. For $s=4$~GeV$^2$, the numerical result, order by order, is now, up to $\mathcal{O}(\alpha_s^5)$
\begin{align}
\delta^{(0)}_{\alpha_s,L\beta_0}(4~{\rm GeV}^2)&=0.09387+0.01372-0.000781-0.00411-0.002062+\cdots\label{eq:delta0-large-beta0}
\end{align}

Returning to the QCD result, with a convenient change to the so-called $C$ scheme for $\alpha_s$~\cite{Boito:2016pwf}, it is possible to obtain the QCD equivalent of Eq.~(\ref{eq:borelrho}). It turns out, as shown in Ref.~\cite{Boito:2020lyp}, that Eq.~(\ref{eq:borelrho}) retains its form in QCD in the $C$ scheme provided an adequate choice of a modified Borel transform is used~\cite{Brown:1992pk}. This means that a similar mechanism of partial cancellation of renormalon singularities is at work in QCD as well, which is another indication that large-$\beta_0$ can be used as a guide to the behavior of the series at higher orders, and lends additional support to the similarities between the series in Eqs.~(\ref{eq:delta0-QCD-2GeV}) and~(\ref{eq:delta0-large-beta0}).

\begin{figure}[!t]
\centering \includegraphics[width=0.5\textwidth]{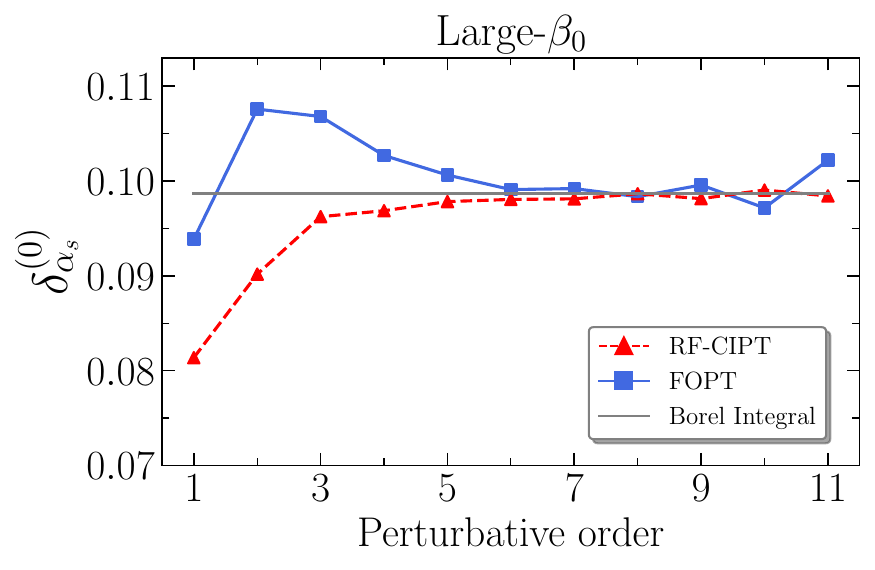}\includegraphics[width=0.5\textwidth]{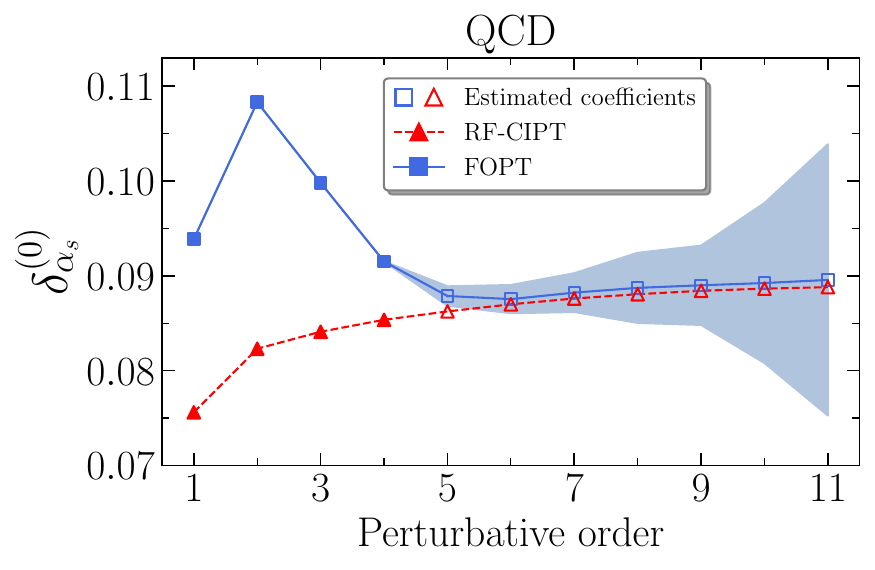}
\caption{Results for $\delta_{\alpha_s}^{(0)}$ in FOPT (blue squares) and in RF-CIPT (red triangles) in the large-$\beta_0$ limit (left-hand panel) and in full QCD (right-hand panel) for $s=(2~{\rm GeV})^2$. The horizontal line in large-$\beta_0$ represents the Borel sum of the series (the imaginary ambiguity is very small in this case and cannot be seen in the plot). In QCD, results beyond $\mathcal{O}(\alpha_s^4)$ employ an estimate for the unknown series coefficients (see text for details). The band in the QCD plot shows the uncertainty due to the unknown coefficients in the FOPT result.}
\label{fig:delta0-RFPT}
\end{figure}

In Fig.~\ref{fig:delta0-RFPT},
we compare the FOPT perturbative series (shown with blue squares) for $\delta^{(0)}_{\alpha_s}$ in the large-$\beta_0$ limit, Eq.~(\ref{eq:delta0-large-beta0}), and in QCD, Eq.~(\ref{eq:delta0-QCD-2GeV}), at $s=(2~{\rm GeV})^2$. In QCD, we use $c_{5,1}=280\pm 140$ and for the higher-order coefficients, $c_{n>5,1}$, we employ the results obtained with Pad\'e approximants in Ref.~\cite{Boito:2018rwt}. The use of other models for the higher-order coefficients~\cite{Beneke:2008ad,Caprini:2019kwp,Benitez-Rathgeb:2022hfj} would lead to similar results. In the large-$\beta_0$ result we see that, in spite of the change in sign at $\mathcal{O}(\alpha_s^3)$, the result does approach the Borel sum and is relatively stable around it at orders 6 to 8. The numerical shift from order 4 to 6 leads to a further correction of $-0.36$\%, which is not very significant given the size of other errors involved in the evaluation of $R_{uds}$. Therefore, although the behavior of this perturbative series is somewhat peculiar, there is no indication of any major convergence issue (in the sense of an asymptotic series), and the result at $\mathcal{O}(a_s^4)$ already provides a good representation for the true value of the series, and even more so the result at $\mathcal{O}(\alpha_s^5)$. The behavior in QCD is very similar, even though the results tend to stabilize at lower values. It is, therefore, reasonable to assume that in QCD the results at order $\mathcal{O}(a_s^4)$ are also approaching the true value from above.

\subsection{Renormalon free scheme}

In order to corroborate the conclusions of the previous section, it is worth investigating another renormalization-scale setting which, effectively, should generate a different asymptotic series to the same true value.

As we have mentioned, an alternate prescription that has been used for a long time in the computation of Adler function moments is CIPT~\cite{LeDiberder:1992jjr,Pivovarov:1991rh}, which consists in setting $\mu^2=-sx$ when using Eq.~(\ref{eq:Adler-expansion}) in the integral of Eq.~(\ref{eq:Adler-mom}). The series thus obtained is no longer a power series in $\alpha_s$, since the running of the coupling is resummed to all orders along the contour of integration. For a long time, the discrepancy between the results obtained in CIPT and those obtained in FOPT was puzzling, especially given that the difference became larger when the $\mathcal{O}(\alpha_s^4)$ term became available, contradicting the expectation that both series were valid asymptotic expansions to the same true value.

This issue is now fully understood. It turns out that CIPT defines a different series, with a different Borel sum, and the expectation that it should agree with FOPT at higher orders was incorrect~\cite{Hoang:2020mkw}. The systematic difference between the two procedures was dubbed `asymptotic separation' and can be calculated exactly if sufficient knowledge is available, such as in the large-$\beta_0$ limit~\cite{Hoang:2020mkw,Hoang:2021nlz,Golterman:2023oml}. This difference is driven by the CIPT sensitivity to IR renormalons: in large-$\beta_0$, more than $99\%$ of this difference is due to the leading IR renormalon, associated with the gluon condensate. This means that, if a consistent procedure is used in order to remove this renormalon, one can then define a modified CIPT which, for all practical purposes, agrees with FOPT. This was demonstrated in Refs.~\cite{Benitez-Rathgeb:2022hfj,Benitez-Rathgeb:2022yqb} where the RF GC scheme was introduced.

In large-$\beta_0$, where exact knowledge about the Borel transform is available, it is possible to implement the RF GC scheme exactly, by the consistent subtraction of the leading IR renormalon. The details of how this is done in practice are summarized in App.~\ref{app:RFGC} but can be found in the original works~\cite{Benitez-Rathgeb:2022hfj,Benitez-Rathgeb:2022yqb}. In the RF scheme, the new CIPT series, to which we refer as `RF-CIPT,' agrees with FOPT at higher orders --- defining two proper asymptotic series to the same true value. The results for RF-CIPT in the large-$\beta_0$ limit are shown in red in the left-hand panel of Fig.~\ref{fig:delta0-RFPT}.\footnote{Here, we use as the IR subtraction scale required for the implementation of the RF GC scheme the value $\eta=0.7s$. Variations of this scale lead to a spread in the values obtained in the RF GC scheme, as discussed in Ref.~\cite{Benitez-Rathgeb:2022yqb}, but this residual scale uncertainty, governed by the renormalization group, is irrelevant to the present discussion.} The series approaches the Borel sum from below, with smaller steps at intermediate orders and without overshooting the true value for low orders. It shows that both series, once properly defined, are fully consistent for sufficiently high orders.

In QCD, the implementation of the RF GC scheme is less simple because the norm of the gluon condensate renormalon is not exactly known. However, as shown in Ref.~\cite{Benitez-Rathgeb:2022hfj}, sufficient knowledge exists to allow a reliable extraction of this norm with sufficient precision for the implementation of the subtracted scheme. The series that is obtained for $\delta^{(0)}_{\alpha_s}$ in RF-CIPT is, order by order, the following
\[
\delta^{(0)}_{\rm RF-CIPT} = 0.07559+0.006716+0.001795+0.001258+0.000884(75)_{c_{51}}.
\]
This series shows a monotonic growth, fixed signs, and a systematic reduction of each consecutive term. It is displayed as the red triangles on the right-hand panel of Fig.~\ref{fig:delta0-RFPT} together with the previous FOPT result. What we see is that the series, although with a rather different behavior, does seem to approach the same value as FOPT at higher orders, in accordance with the expectation of the RF GC scheme. What is more, at $\mathcal{O}(\alpha_s^5)$ the two asymptotic series are already very similar, which corroborates that no big surprise should be expected for even higher orders.
The results in QCD are very similar, qualitatively, to those obtained in the large-$\beta_0$ limit, where the Borel sum is known. This further corroborates that the true value of the series is very likely well approximated by the results at $\mathcal{O}(\alpha_s^5)$.

\section{Other corrections to \boldmath \texorpdfstring{$R_{uds}(s)$}{Ruds}}
\label{sec:othercorrections}

\subsection{Duality violations}

The use of pQCD to describe $R(s)$ is predicated on the assumption of quark-hadron duality. For sufficiently large $s$ and far away from resonances one expects that pQCD should provide a good approximation to $R_{uds}$. At intermediate values of $s$, residual oscillations due to superimposing tails of higher resonances can still be present, and represent an intrinsic limitation of pQCD in the Minkowski. These damped oscillations around pQCD are what we refer to as the DVs.

Their description, however, cannot be obtained solely from first principles, and one must rely on assumptions about the QCD spectrum in order to derive a reasonable parametrization for the DV contribution. The DV contribution is encoded in the function $\Delta(z)$, which represents the deviations of $\Pi(z)$ from the OPE description
\beq
\Pi(z) = \Pi_{\rm OPE}(z) + [\Pi(z)-\Pi_{\rm OPE}(z)] = \Pi_{\rm OPE}(z) +\Delta(z).
\eeq
Then, the DV contribution to $\rho_{\rm EM}(s)$ is simply $\rho_{\rm EM}^{\rm DVs}(s)=(1/\pi){\rm Im} \Delta(s)$.
A formulation of the problem in terms of the Borel-Laplace transform of $\Pi(q^2)$ and hyperasymptotics was presented in Ref.~\cite{Boito:2017cnp}. With the assumption that the QCD spectrum is Regge-like for high energies and in the $N_c\to \infty$ limit, one can derive the following parametrization valid for large $s$
\beq
\rho^{\rm DVs}(s) = e^{-\delta-\gamma s}\sin(\alpha+\beta s).\label{eq:DV-parametrization}
\eeq
The general form of the sub-leading corrections to this parametrization can be found in Ref.~\cite{Boito:2017cnp} and were investigated recently~\cite{Boito:2025pwg,Boito:2024gtb}, but these corrections have been shown to be small, in the case of $1/s$ corrections, or too slowly varying in $s$, in the case of $1/\log s$ corrections, to have an impact in the description of experimental data. We will therefore use the parametrization as given in Eq.~(\ref{eq:DV-parametrization}).

Since DVs are related to resonances their quantitative description is channel specific. In hadronic $\tau$ decays, e.g., where they have been studied for a number of years~\cite{Boito:2011qt,Boito:2020xli,Boito:2025pwg,Rodriguez-Sanchez:2025nsm}, the vector-isovector and axial-vector-isovector contributions intervene. In $R(s)$, one must consider the vector-isoscalar contribution as well. The inclusion of the $I=0$ channel is not so straightforward because the parametrization of Eq.~(\ref{eq:DV-parametrization}) was obtained in the chiral limit, and it is not clear how to systematically include $m_s$ effects. Here we follow the strategy of Ref.~\cite{Boito:2018yvl}. Ignoring the small disconnected contribution and taking into account the quark charges one arrives at the following parametrization for the DVs in the EM spectral function
\beq
\rho_{\rm EM}^{\rm DVs}(s) = \frac{5}{9}e^{-\delta_1-\gamma_1 s}\sin(\alpha_1+\beta_1 s)+\frac{1}{9}e^{-\delta_0-\gamma_0 s}\sin(\alpha_0+\beta_0 s).\label{eq:rhoDVEM}
\eeq
In this form, the term proportional to 5/9, corresponds, in the isospin limit, to the vector-isovector DV contribution that appears in hadronic $\tau$ decays. Given the observation that the $\rho$ meson spectrum is nearly degenerate with the $\omega$ spectrum one expects that the DVs in $I=1$ and $I=0$ channels are degenerate in shape for large $s$. Therefore, an additional assumption was made in Ref.~\cite{Boito:2018yvl}, which consists in assuming that $\beta_0=\beta_1$ and $\gamma_0=\gamma_1$. With this assumption, that we adopt here as well, the vector-isoscalar DV contribution contains only two new parameters with respect to the vector-isovector counterpart, namely $\delta_0$ and $\alpha_0$.

Since the parametrization of Eq.~(\ref{eq:rhoDVEM}) is valid only for `large $s$', the use of this parametrization is predicated on the assumption that the large $s$ regime is already attained. This assumption is supported by data since the $\tau$-based $I=1$ vector analyses~\cite{Boito:2025pwg,Boito:2020xli} show that good fits can be obtained for $s\gtrsim 1.4$~{GeV}$^2$. For the $I=0$ part, the analyses of Ref.~\cite{Boito:2018yvl} support the notion that this limit is reached for $s\gtrsim 2.9$~GeV$^2$. Therefore, for the description of the inclusive $R(s)$ data, where $s>3.24$~GeV$^2$, our parametrization is expected to be valid as well.

The DV contribution to $R_{uds}(s)$ parametrized as in Eq.~(\ref{eq:RudsNLO}) is finally written as
\beq
\delta_{\rm DVs} = 6\pi^2 \rho_{\rm EM}^{\rm DVs}(s),\label{eq:deltaDVs}
\eeq
where $\rho_{\rm EM}^{\rm DVs}(s)$ is given in Eq.~(\ref{eq:rhoDVEM}).

Here, the vector-isovector DV parameters, $\delta_1$, $\gamma_1$, $\alpha_1$, and $\beta_1$, will be fixed using results from a recent hadronic $\tau$ decay analysis~\cite{Boito:2025pwg}, in which an improved inclusive vector-isovector spectral function was built from the combination of the available $\tau \to 2\pi\nu_\tau$, $\tau\to 4\pi\nu_\tau$, and $\tau \to \bar K K \pi$ data supplemented with smaller contributions from higher multiplicity channels obtained from $e^+e^-\to {\rm hadrons}$ cross-sections using CVC~\cite{Boito:2020xli}. In particular, we use the parameters from the fit of Table 2 of Ref.~\cite{Boito:2025pwg} with $s_0^{\rm min}=1.5747~{\rm GeV}^2$ which read
\begin{align}
\delta_1&= 3.01(39), \\
\gamma_1&= 0.87(24)~{\rm GeV}^{-2},\\
\alpha_1&= -1.34(73),\\
\beta_1&= 3.78(38)~{\rm GeV}^{-2}.
\end{align}
Their full covariance matrix is used in error propagations.

To fix the two additional isoscalar parameters, $\delta_0$ and $\alpha_0$, we employ the strategy of Ref.~\cite{Boito:2018yvl} which consists in fitting the $R_{uds}$ data obtained from the measurements of {\it exclusive} channels in the interval $ 3.3~{\rm GeV}^2 \leq s\leq 4~{\rm GeV}^2$. We use the data from the combination of exclusive-channel cross-sections of Ref.~\cite{Keshavarzi:2018mgv}, to which we refer as `KNT.' With the $I=1$ parameters quoted above, we obtain
\begin{align}
\delta_0&=0.96(22) ,\\
\alpha_0&= 0.80(27).
\end{align}
These values are similar to those quoted in Ref.~\cite{Boito:2018yvl} but can be considered as an update of those numbers given the new knowledge about the $I=1$ parameters. The correlation of $45\%$ among these values is considered in error propagations.

With this set of parameters, the DV contribution to $R_{uds}$ in the inclusive region is {\it fixed} and, for $s\geq 4~{\rm GeV}^2$, is an extrapolation of results obtained at lower energies and from fits to other data sets (hadronic $\tau$ decays~\cite{Boito:2020xli} and exclusive $R_{uds}(s)$ data~\cite{Keshavarzi:2018mgv,Boito:2018yvl}). We have also tried fits of $I=0$ parameters to the inclusive $R_{uds}$ data but we obtain results that are very similar to those based on the extrapolations, with no significant difference. Therefore, we prefer to use the description based on the extrapolation as our central value, since in this case the DVs are fixed by external information, which is arguably more interesting.

\subsection{Quark-mass corrections}

Beyond perturbation theory in the chiral limit, one should consider the quark-mass corrections. These are known up to $\alpha_s^3$~\cite{Chetyrkin:1993hi, Chetyrkin:1990kr,Gorishnii:1986pz}. Corrections from $m_u$ and $m_d$ can safely be neglected and, for $R_{uds}$, one can consider only those from $m_s$. Since the contribution is numerically small, with an overall scale at $s=(4~{\rm GeV^2})$ set by
\[
2a_s \frac{m_s^2(s)}{s} \sim 4\times 10^{-4},
\]
where we used the $\MSbar$ value $m_s(2~{\rm GeV})=93.5(8)$~MeV~\cite{ParticleDataGroup:2024cfk}, we consider these corrections up to $\mathcal{O}(\alpha_s^2)$, as done, e.g. in Refs.~\cite{Boito:2022rkw,Davier:2019can}. The result for $\delta_{m_q^2}$, defined in Eq.~({\ref{eq:RudsNLO}}), is then
\beq
\delta_{m_q^2} = \frac{m_s^2(s)}{s}\left(2a_s+\frac{227}{12}a_s^2+\cdots \right).\label{eq:deltams}
\eeq
When estimating the uncertainty on this contribution we consider the uncertainty from the value of $m_s(2~{\rm GeV})$ as well as that from $\alpha_s$.
We remark, finally, that the $\mathcal{O}(a_s^2)$ correction is of the same order of the leading correction, but the overall contribution remains very small.

\subsection{Electromagnetic corrections}

Because the $R(s)$ data include EM corrections we have to consider these in our description of $R_{uds}(s)$. The only numerically important contribution arises from one-photon exchange~\cite{Yndurain:1977xb,Kataev:1992dg,Surguladze:1996hx} and leads, in the decomposition of Eq.~(\ref{eq:RudsNLO}), to~\cite{Boito:2018yvl}
\beq
\delta_{\rm EM} = \frac{\alpha_{\rm EM}}{4\pi},\label{eq:deltaEM}
\eeq
where we use $\alpha_{\rm EM} =1/137.0356$~\cite{ParticleDataGroup:2024cfk}.
Since this contribution is numerically very small, its uncertainty can safely be neglected.

\section{Results for \boldmath \texorpdfstring{$R_{uds}$}{Ruds}}
\label{sec:R-results}

With the contributions from pQCD in the FOPT prescription, Eq.~(\ref{eq:delta0-QCD-2GeV}), the contributions from DVs of Eq.~(\ref{eq:deltaDVs}), the strange-mass corrections, Eq.~(\ref{eq:deltams}), and the EM correction of Eq.~(\ref{eq:deltaEM}), we can obtain the complete results for $R_{uds}$ as parametrized in Eq.~(\ref{eq:RudsNLO}). In Tab.~\ref{tab:Ruds_contributions} we give representative values of $R_{uds}$ in the inclusive region with the breakdown into the different corrections. The contribution from pQCD is always the dominant one, but for lower energies in the inclusive region, DVs can also be relevant. For $\sqrt{s}=2~{\rm GeV}$, e.g., pQCD represents a $\sim 9\%$ correction, with a negative $\sim 3\%$ correction from the DV contribution, followed by a $\sim 0.3\%$ $m_s$ correction and a much smaller contribution from EM. At $\sqrt{s} =2.5~{\rm GeV}$ the DV contribution is at the sub-percent level, while pQCD is still $\sim 8\%$, and for $\sqrt{s} \geq 3.0~{\rm GeV}$ the DV contribution is already insignificant. The DV parameters, however, are not well controlled given the present knowledge; the error associated with the DVs is large and, at lower energies, strongly dominates the total error of $R_{uds}$. In Fig.~\ref{fig:R-theory-and-data} we can see that, in spite of the large errors, the central value from the description that includes the DVs (dashed blue line) reproduces remarkably well the oscillations seen in KEDR data. We remind that this result is {\it not a fit} to the data shown in Fig.~\ref{fig:R-theory-and-data}, it is, rather, an extrapolation from the description with $I=1$ parameters fixed by $\tau$ decay data~\cite{Boito:2025pwg} and $I=0$ parameters fixed by exclusive region $R_{uds}$ data~\cite{Boito:2018yvl,Keshavarzi:2018mgv}.

\begin{table}[!t]
\begin{center}
\caption{Results for $R_{uds}$ in the inclusive region and the respective breakdown into the different corrections defined in Eq.~(\ref{eq:RudsNLO}). }
\begin{tabular}{clllll}
\toprule
$\sqrt{s}~(\text{GeV})$ & $R_{uds}$ & $\delta^{(0)}_{\alpha_s}$ & $\delta_{\rm DVs}$ & $\delta_{m_q^2}$&$\delta_{\rm EM}$ \\ \midrule
$2.0$ & $2.12(14)$ & $0.0879(21)$ & $-0.030(69)$ & $0.000775(27)$ & $0.00058$\\
[0.1cm]
$2.5$ & $2.178(19)$ & $0.0822(15)$ & $0.0060(93)$ & $0.000377(11)$ & $0.00058$ \\
[0.1cm]
$3.0$ & $2.1560(46)$ & $0.0776(12)$ & $-0.0004(19)$ & $0.0002148(54)$ & $0.00058$ \\
[0.1cm]
$3.5$ & $2.1491(22)$ & $0.0739(11)$ & $-0.00003(13)$ & $0.000136(31)$ & $0.00058$ \\ \bottomrule
\label{tab:Ruds_contributions}
\end{tabular}
\end{center}
\end{table}

\section{Comparison between theory and data}
\label{sec:Comparison}

\begin{figure}[!t]
\centering \includegraphics[width=0.9\textwidth]{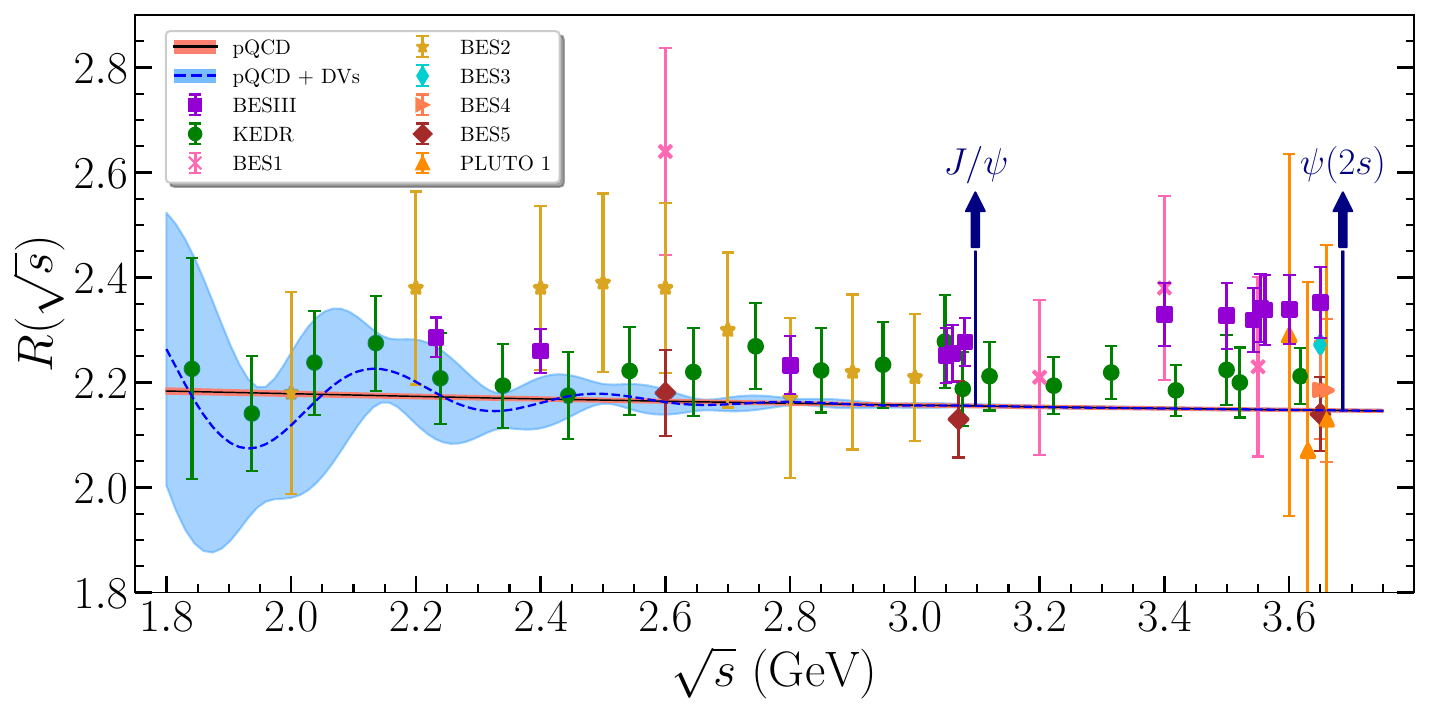}
\caption{Inclusive data for $R_{uds}$ from several experiments compared with pQCD (black line) and pQCD$+$DVs (blue-dashed line). The description including DVs is {\it not a fit} to the data shown in the figure and is obtained extrapolating from hadronic $\tau$-decay and exclusive $R_{uds}$ data analyses~\cite{Boito:2025pwg,Boito:2018yvl,Keshavarzi:2018mgv}. Both curves include the small strange-quark mass and EM corrections. }
\label{fig:R-theory-and-data}
\end{figure}

We turn now to a quantitative comparison between theory and the available experimental inclusive data for $R_{uds}$. As already shown in Fig.~\ref{fig:R-theory-and-data}, two recent data sets play a prominent role in this comparison, namely the BES-III~\cite{BESIII:2021wib} and the KEDR~\cite{KEDR:2018hhr,Anashin:2015woa,Anashin:2016hmv} analyses. The BES-III measurements are the most precise, but not many data points are available below 3.4~GeV, while KEDR, albeit with larger errors, provide data in the whole interval of Fig.~\ref{fig:R-theory-and-data}. In addition to these more recent and precise measurements, we consider BES1~\cite{BES:1999wbx}, BES2~\cite{BES:2001ckj}, BES3~\cite{BES:2006dso}, BES4~\cite{BES:2006pcm}, and BES5~\cite{BES:2009ejh} data sets (to which we sometimes refer collectively as `BES') and the PLUTO collaboration results of Ref.~\cite{Criegee:1981qx}, as these remain relevant within the considered energy region.
For the BES-III data set, a non-trivial covariance matrix can be built from the information given in the original publication, while the KEDR experiment gives their covariance matrix explicitly.
In cases where no information on non-trivial correlations is provided, we assume fully correlated systematic uncertainties and uncorrelated statistical ones.

\begin{figure}[!t]
\centering \includegraphics[height=0.6\textwidth]{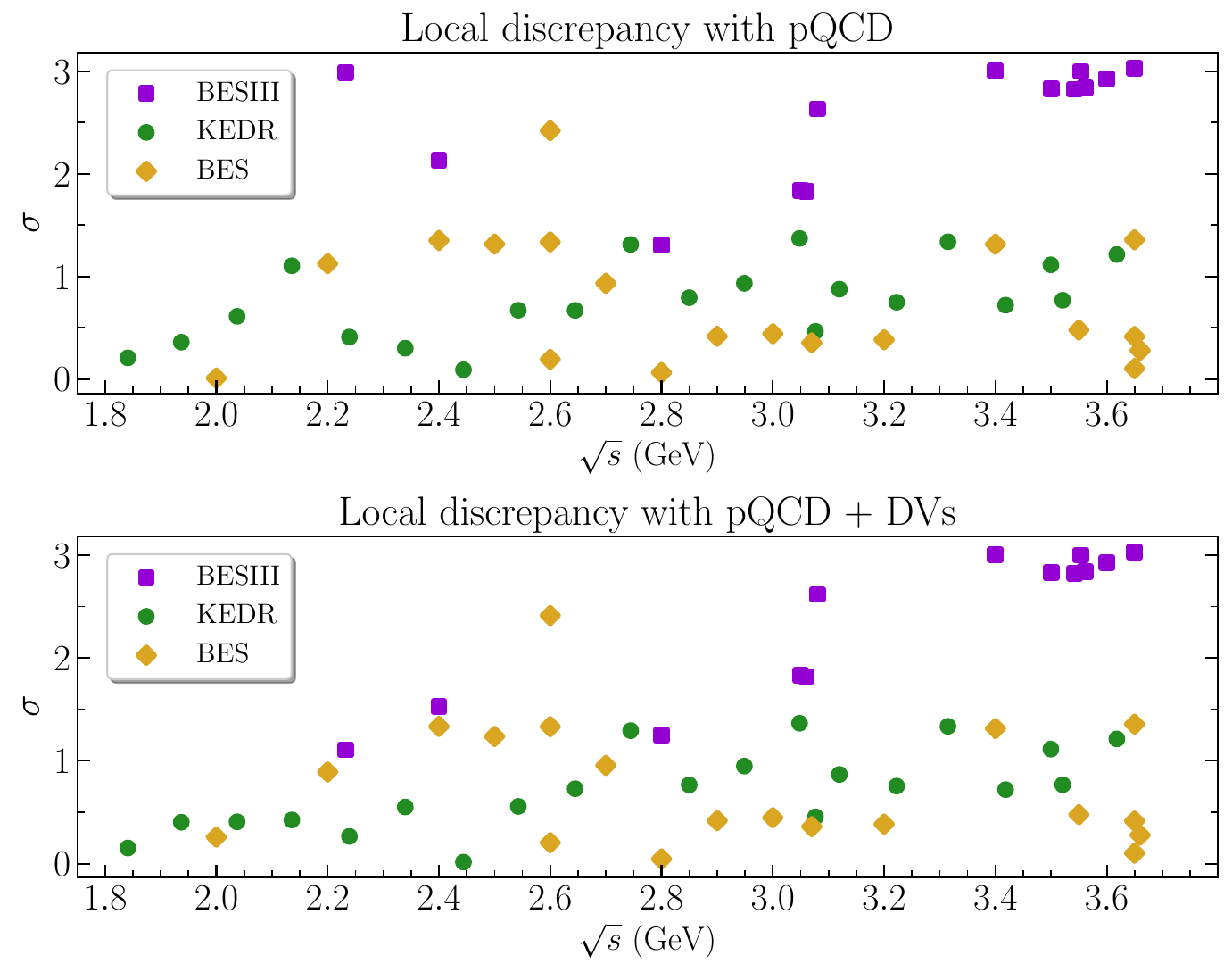}
\caption{Local discrepancies of data points from BES-III, KEDR, and BES with respect to pure pQCD (upper panel) and pQCD+DVs (lower panel). In both cases small $m_s$ and EM corrections are taken into account.}
\label{fig:discrepancies_R}
\end{figure}

We start by investigating the local discrepancies between the experimental data points and the theory description of $R_{uds}$.
At lower energies, as shown in Fig.~\ref{fig:R-theory-and-data},
KEDR results show signs of an oscillation around perturbation theory, while they are systematically above it for larger energies below charm threshold. The BES-III data, on the other hand, are always systematically above pQCD, with the data points with $\sqrt{s}\geq 3.4$~GeV being particularly discrepant with theory predictions. In Fig.~\ref{fig:discrepancies_R}, we show the significance of the discrepancy with respect to theory for the KEDR, BES-III, and all BES data sets, point by point; excluding the last BES-III data point at 3.671~GeV since it may be too close to the $\psi(3770)$ resonance.\footnote{Since the $J/\psi$ and the $\psi(2S)$ are extremely narrow~\cite{ParticleDataGroup:2024cfk}, we do not exclude points on the basis of their proximity to these narrow charm resonances.}
The KEDR data, although systematically larger than pure pQCD for most of the interval, is in agreement with theory within less than 2$\sigma$. The inclusion of the DV contribution reduces even more the discrepancy, not only because the central values are in better agreement but, more importantly, because the theory errors become much larger. Several of the BES-III data points have a tension with pure pQCD above 2$\sigma$, with 8 points showing $\sim 3\sigma$ discrepancies with theory, especially above $3.4$~GeV. One should note, however, that the BES-III data points at these higher energies are highly positively correlated ($>86\%$) and less information is available than Fig.~{\ref{fig:R-theory-and-data}} seems to indicate. The inclusion of DVs in the theory description significantly reduces the discrepancy for the first two data points, at lower energies, but has essentially no effect for the rest of the data. The BES data points, in turn, are, in general, compatible with theory within less than 2$\sigma$, although the BES2 data set also tend to be systematically above pQCD.

In the spirit of data-driven $a_\mu^{\rm HVP}$ analyses, it is interesting to combine the available experimental data in a single data set, which condenses the available information and allows for a more quantitative assessment of the compatibility between the different measurements. Here, we combine the data of Fig.~\ref{fig:R-theory-and-data}
using the KNT algorithm of Refs.~\cite{Keshavarzi:2018mgv,Keshavarzi:2019abf}. We stress, however, that, contrary to the KNT analyses, the scope of our data combination is more modest as it is restricted to the energy interval of Fig.~\ref{fig:R-theory-and-data}, with the aim of investigating $R_{uds}$, and is not influenced by the data points that lie outside this energy region, through, for example, long-distance correlations.

Here, we describe the KNT algorithm succinctly, and we refer to the original works~\cite{Keshavarzi:2018mgv,Keshavarzi:2019abf} and related applications of the same algorithm~\cite{Boito:2025pwg} for a detailed description. In a nut-shell, the KNT data combination algorithm starts with the definition of the number of {\it clusters}, $N_{\rm cl}$. To each cluster $m$, with $m=1,...,N_{\rm cl}$, one assigns consecutive data points from
different experiments. The cluster sizes can vary along the spectrum, but one must avoid over-
and under-populated clusters. The final set of $N_{\rm cl}$ combined $R^{(m)}$ data points sit at the energies $E^{(m)}$ which are referred to as {\it cluster centers}, determined by the weighted average of the energy values of each data point belonging to cluster $m$. The fit procedure is initialized with $R^{(m)}$ values that are simply weighted averages of the $R(s)$ values of data points inside cluster $m$. A representation of $R_{uds}$, denoted $\mathcal{R}(s;R^{(m)})$, is obtained from the $R^{(m)}$ values by a piece-wise linear interpolation between consecutive $R^{(m)}$ values (extrapolation is used for data points that lie outside the interval defined by the cluster centers). The final results are obtained from a $\chi^2$ fit of $\mathcal{R}(s;R^{(m)})$ to all of the experimental data points $d_i$, with as free parameters the $N_{\rm cl}$ values of $R^{(m)}$. This $\chi^2$ function can then be written as
\begin{equation}
\chi^2(R^{(m)})=\sum_{i=1}^N\sum_{j=1}^N\left(d_i-\mathcal{R}(s_i; R^{(m)})\right)(C^{-1})_{ij}\left(d_j-\mathcal{R}(s_j; R^{(m)})\right),
\end{equation}
where $N$ is the total number of data points and $C$ is the covariance-matrix of the data points. For a linear piece-wise $\mathcal{R}(s;R^{(m)})$, this $\chi^2$ can be minimized analytically.\footnote{We have checked that using quadratic interpolation between the $R^{(m)}$ data points we obtain essentially the same results.}

An interesting by-product of this type of data combination is that one can quantify the compatibility between different experimental data points by inspecting the $\chi^2$ contribution and the respective $p$-value from individual clusters, $\chi^2_{m}$ and $p_{m}$-values, respectively. Clusters with small $p_{m}$-values contain conflicting data points. One can account for the tension between data points in clusters with low $p_{m}$-values by applying an error inflation procedure. Here, when $\chi^2_{m}/({\rm d.o.f.})_m>1$ we rescale the final error in $R^{(m)}$ by a factor $\sqrt{\chi^2_m/({\rm d.o.f.})_m}$~\cite{Keshavarzi:2018mgv}.

\begin{figure}[!t]
\centering \includegraphics[height=0.5\textwidth]{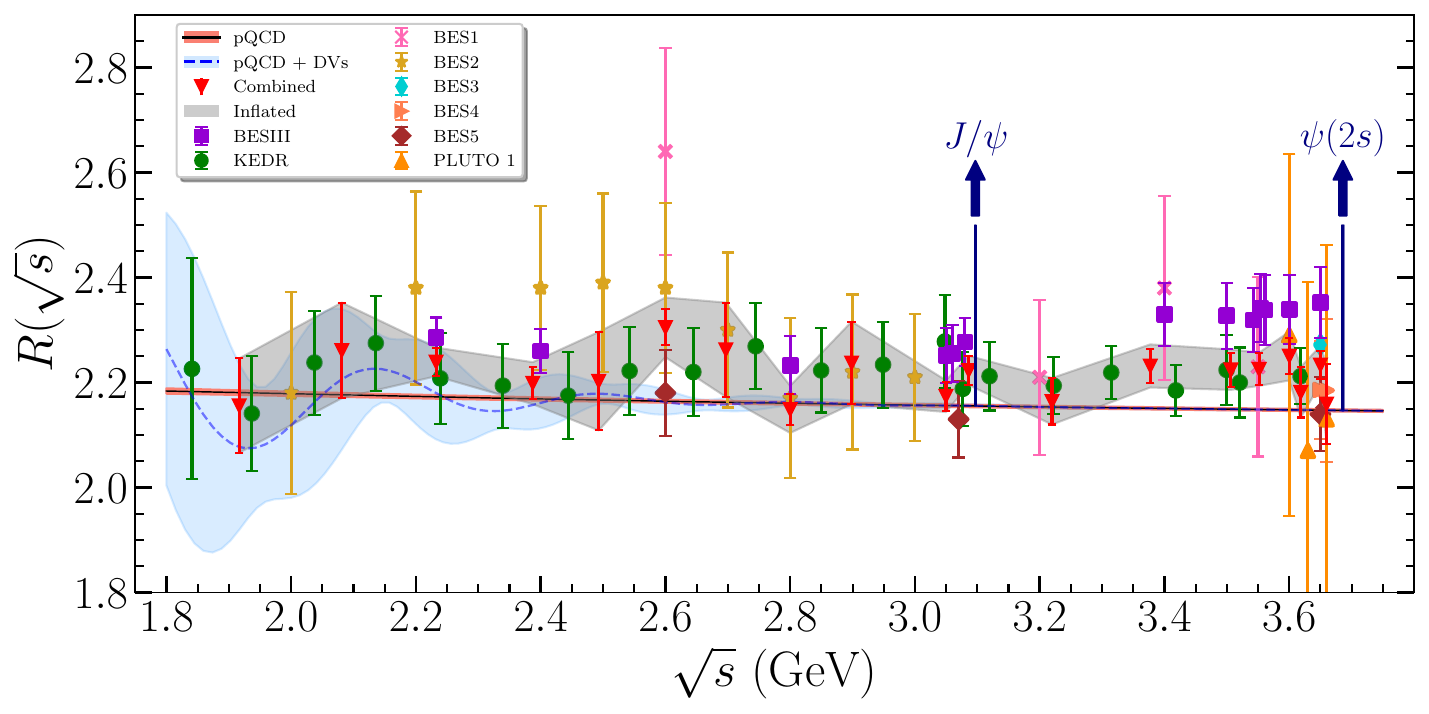}
\caption{Data combination (red triangles) with 19 clusters using the algorithm discussed in the text. The red error bars show errors before error inflation. The gray band gives the errors after error inflation is applied to account for local tensions between data points in the same cluster (the effect is small).}
\label{fig:combined_R}
\end{figure}

\begin{figure}[!t]
\centering \includegraphics[height=0.35\textwidth]{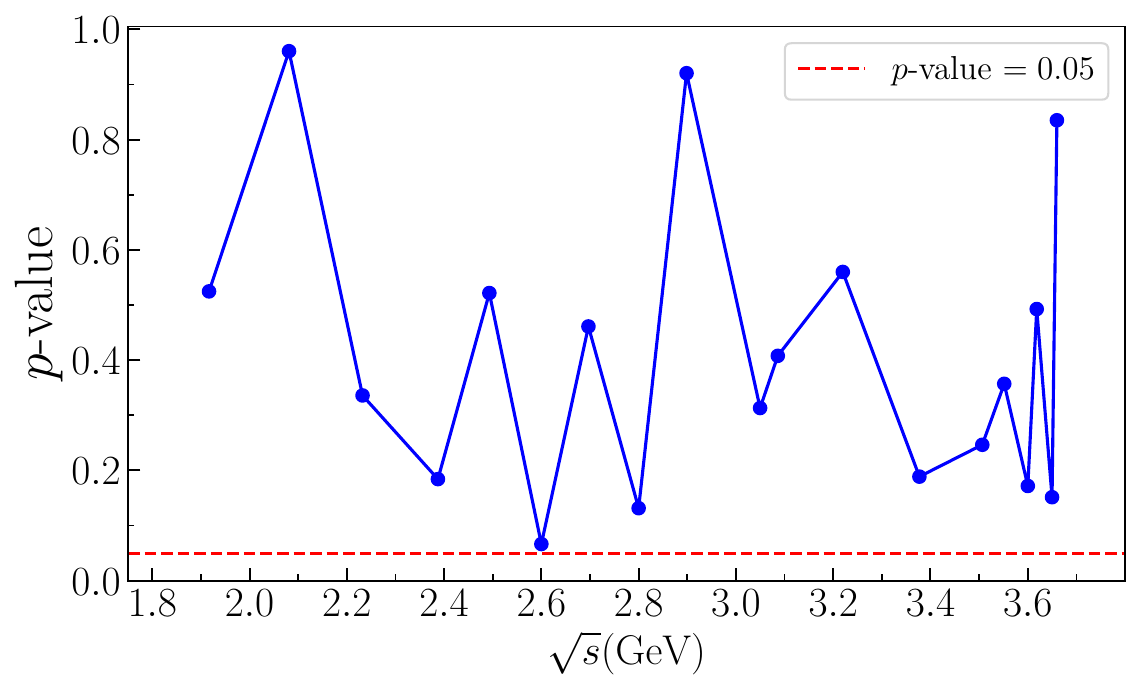}
\caption{Local $p$-value for each of the 19 clusters of the data combination shown in Fig.~\ref{fig:combined_R}. The red-dashed line shows the value 5\% to guide the eye. }
\label{fig:local_pvalues}
\end{figure}

A combination as described above of the available data between 1.8~GeV and 3.66~GeV (56 data points) with 19 clusters is shown in Fig.~\ref{fig:combined_R} (red-triangles and gray band). This combination has a $\chi^2/{\rm d.o.f.} = 41.67/37=1.13$ with an associated global $p$-value of $27.5\%$. This shows that the data are, on average, compatible. An inspection of the $p$-value per cluster, shown in Fig.~\ref{fig:local_pvalues}, reveals that the smallest is $6.6\%$, which is not alarming. Therefore, local discrepancies in each cluster are also relatively mild, in spite of the visible tensions that exist in the data sets. Locally, the combined results differ from the theory prediction by at most 2.4$\sigma$, which happens for the cluster with $\sqrt{s}=2.6$~GeV.

\begin{figure}[!t]
\centering \includegraphics[width=0.35\textwidth]{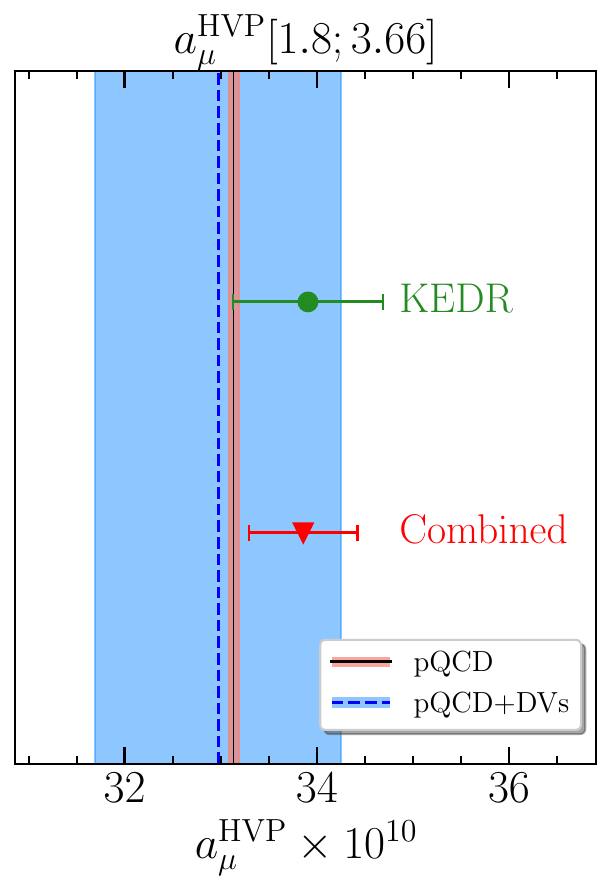}\includegraphics[width=0.35\textwidth]{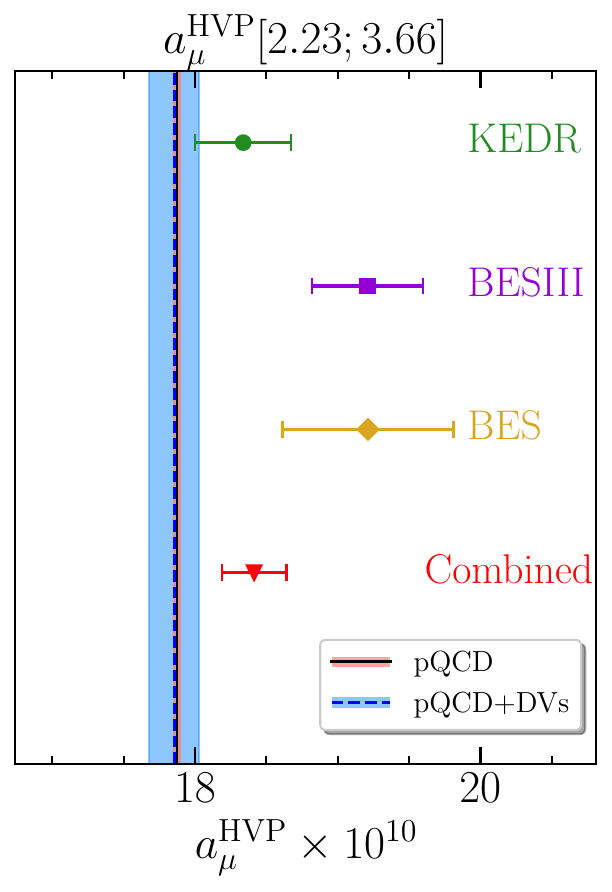}
\caption{Results for $\amuHVP$ in two energy intervals from theory and data. Left panel: $1.8~{\rm GeV}\leq\sqrt{s} \leq 3.66~{\rm GeV}$, results from KEDR and the combined data compared with pure pQCD and pQCD + DVs (always including small $m_s$ and EM corrections). Right panel: $2.23~{\rm GeV}\leq\sqrt{s} \leq 3.66~{\rm GeV}$, results from KEDR~\cite{KEDR:2018hhr,Anashin:2015woa,Anashin:2016hmv}, BES-III~\cite{BESIII:2021wib}, BES~\cite{BES:1999wbx, BES:2001ckj, BES:2006dso, BES:2006pcm, BES:2009ejh}, and the combined data compared with pure pQCD and pQCD+DVs. }
\label{fig:amu_with_discrepancies}
\end{figure}

We turn now to a comparison of integrated values for $a_\mu^{\rm HVP}$ in two different energy intervals. First, we consider the full range of inclusive $R_{uds}$ data below charm, i.e. $1.8$~GeV to $3.66$~GeV (again excluding the last BES-III data point, with bin center at 3.671~GeV). In this range, we compute $a_{\mu}^{\rm HVP}[1.8;3.66]$ from KEDR data and from the combined data set and compare them with the theory predictions from pure pQCD and pQCD + DVs.
The results appear in the second row of Tab.~\ref{tab:amu_values} and on the left-hand panel of Fig.~\ref{fig:amu_with_discrepancies}. Results from KEDR and the combined data are larger than the theory prediction by about $0.8\times 10^{-10}$ but agree with theory within $0.6\sigma$ for pQCD$+$DVs and within a little over $1\sigma$ for pure pQCD. The second energy interval that we investigate is the interval for which we have the BES-III data set, $2.23$~GeV to $3.66$~GeV. In this interval we can also compute $a_{\mu}^{\rm HVP}[2.23;3.66]$ from KEDR, the BES data sets, and the combined data.\footnote{To integrate all data sets in the same interval, we interpolate between data points, when necessary, taking into account all correlations.} Results for $\amuHVP$ are again given in Tab.~\ref{tab:amu_values} and Fig.~\ref{fig:amu_with_discrepancies}, while the discrepancies with respect to theory are quantified in Tab.~\ref{tab:amu_discrepancies}. The BES-III result differs by $3.4\sigma$ with respect to pure pQCD result, and by $3.2\sigma$ with respect to the pQCD $+$ DVs result. Results from the BES data sets show a tension of $2.2\sigma$, while KEDR are compatible with either of the theory descriptions within $\sim 1.4\sigma$. The result for $a_{\mu}^{\rm HVP}[2.23;3.66]$ from the combined data are $2.4\sigma$ apart from pure pQCD and $2.0\sigma$ from pQCD+DVs. Although the discrepancy for $a_{\mu}^{\rm HVP}[2.23;3.66]$ from BES-III and pQCD is larger than $3\sigma$, the result obtained from the BES-III data is only $1.4\times 10^{-10}$ larger, which is not a large discrepancy in absolute terms for the total $a_\mu^{\rm HVP}$, given the present uncertainties and the much larger discrepancies that appear in the contribution from the $\rho$-meson peak~\cite{Aliberti:2025beg}.

\section{Conclusions}
\label{sec:Conclusions}
In this paper, we have investigated the theory description of $R(s)$ below open-charm threshold within a framework that includes pQCD and DV contributions, as well as small quark-mass and EM corrections. Discrepancies between theory and data for $R_{uds}$ have conceptual implications for the validity of pQCD, and can impact the SM assessment of $g-2$ of the muon, as well as determinations of $\alpha_s$ and the charm-quark mass.

\begin{table}[!t]
\begin{center}
{\small
\caption{Results for $a_\mu^{\rm HVP}\times 10^{10}$ for two energy intervals: $1.8~{\rm GeV}\leq\sqrt{s} \leq 3.66~{\rm GeV}$ and $2.23~{\rm GeV}\leq\sqrt{s} \leq 3.66~{\rm GeV}$. We give theory results from pure pQCD and pQCD $+$ DVs (in both cases including small $m_s$ and EM corrections), as well as results obtained from the experimental data from BES-III~\cite{BESIII:2021wib}, KEDR~\cite{KEDR:2018hhr,Anashin:2015woa,Anashin:2016hmv}, and BES~\cite{BES:1999wbx, BES:2001ckj, BES:2006dso, BES:2006pcm, BES:2009ejh}. The result from the data combination (`Comb.') of Fig.~\ref{fig:combined_R} is shown in the last column.}
\begin{tabular}{ccccccc}
\toprule
$\sqrt{s}$~[GeV] & pQCD & pQCD$+$ DVs & BESIII & KEDR& BES & Comb. \\
\midrule
$[1.8;3.66]$ & $33.087(51)$ &$32.9(1.3)$ & --- & $33.91(78)$ & --- & $33.86(56)$\\
[0.1cm]
$[2.23;3.66]$ &$17.858(23)$ &$17.84(18)$ & $19.21(39)$ & $18.34(34)$ & $19.21(60)$ & $18.42(23)$ \\
\bottomrule
\label{tab:amu_values}
\end{tabular}}
\end{center}
\end{table}

First, we have shown that the pQCD $\alpha_s$-expansion obtained in standard FOPT has a somewhat peculiar behavior: it apparently overshoots the true value in the first few orders and only later approaches a more stable result at higher orders at the expense of a change in sign for the coefficients. We used the large-$\beta_0$ limit and a model for pQCD higher-order coefficients to corroborate the conclusion that this series, albeit not completely `standard,' does not show signs of a problem and that the results at $\mathcal{O}(\alpha_s^4)$, known exactly~\cite{Baikov:2008jh,Herzog:2017dtz}, or order $\mathcal{O}(\alpha_s^5)$, which can be estimated, are, very likely, reasonable approximations to the true value of the series given our present precision. This conclusion was further corroborated with the use of the RF GC scheme~\cite{Benitez-Rathgeb:2022yqb,Benitez-Rathgeb:2022hfj}, designed to eliminate the leading IR renormalon of the pQCD series. The RF-CIPT series approaches the true value in the large-$\beta_0$ limit monotonically from below, and displays a very similar behavior in pQCD when we extend the series with Pad\'e-approximant results for higher orders. This again reinforces that pQCD is under control and is not likely to change significantly with unknown higher-order terms. We have performed a careful estimate of the error associated with the truncation of perturbation theory, and this error is very small when compared with the experimental uncertainties. In conclusion, there is very little room for changes in the pQCD contribution.

We then investigated quantitatively the DV contribution, which account for the exponentially-damped residual oscillations in the spectral function. Our description of DVs is based on previous works~\cite{Boito:2017cnp,Cata:2005zj,Cata:2008ye}. The free DV parameters were fixed with external information from a hadronic $\tau$-decay analysis~\cite{Boito:2025pwg} and an analysis of exclusive KNT $R(s)$ data below $2.0$~GeV~\cite{Boito:2018yvl}. The results obtained from this pQCD $+$ DVs description nicely follow the oscillations observed in the KEDR data set, seen in Fig.~\ref{fig:R-theory-and-data}. Because the DV parameters are not well known, especially $\gamma_{1,0}$ and $\beta_{1,0}$, the error on the DV contribution is still very large. Although very visible for the lower-energy part of the inclusive $R_{uds}$ data, the DV contribution is essentially insignificant for $\sqrt{s}>2.8$~GeV, and can safely be neglected. These conclusions are predicated on the assumptions that underlie the DV parametrization that we employ, especially the assumption that the large $s$ regime is already attained. But one can also see the nice agreement of the central values with respect to KEDR data points as a test of these assumptions --- a test that the description passes successfully, although one should bear in mind that errors are large.

We have then shown that the KEDR data are in good agreement with pQCD or the pQCD $+$ DV description of the $R_{uds}$ data, albeit the KEDR data tend to be systematically larger than theory.
The BES-III data, on the other hand, are always larger than pQCD and many data points show a tension of the order of $3\sigma$ with pQCD results, especially for the set of data points with $\sqrt{s}\geq 3.4$~GeV. The inclusion of DVs improves the agreement between data and theory only for the first two data points, with $\sqrt{s}\leq 2.4$~GeV. The BES data sets also have a tendency to be larger than pQCD, but agree with theory within $2\sigma$ for most of the points.

We have also performed a combination of the available experimental data using the algorithm of Refs.~\cite{Keshavarzi:2018mgv,Keshavarzi:2019abf}. The data combination shows that the different data sets are statistically compatible, in spite of a few local tensions that we account for by applying an error inflation procedure. The combined data are compatible with theory within at most $2.4\sigma$.

Results for the integrated contributions to $a_\mu^{\rm HVP}$ in two different intervals show that the KEDR data lead to results fully compatible with pQCD and pQCD+DVs. Results from the BES experiments are larger than the theory counterparts but are also compatible within 2.2$\sigma$. The combined data, with smaller errors, also lead to a $a_\mu^{\rm HVP}[2.23;3.66]$ marginally compatible with pQCD (2.4$\sigma$) and compatible with the result that includes the DV contribution. Finally, the BES-III data lead to a result for $a_\mu^{\rm HVP}[2.23;3.66]$ more than $3\sigma$ away from pQCD.

In conclusion, below $2.5$~GeV there is room for a potentially sizable DV contribution, which could soften some of the discrepancies between theory and data points that are observed in $R_{uds}$. This is not the case for the data with $\sqrt{s}>2.8$~GeV, where the DVs are tiny. For these energies, there is very little room for change in the theory description and any incompatibility between data and theory is disconcerting. In particular, we cannot find any mechanism to account for the larger values obtained by the BES-III collaboration for $\sqrt{s}\geq 3.4$~GeV. We hope the present work will encourage further experimental investigations of $R_{uds}$ in this energy range.

\begin{table}[!t]
\centering
\small{\caption{Discrepancies between the results for $\amuHVP$ from theory and data (see Fig.~\ref{fig:amu_with_discrepancies}) in two energy intervals $1.8~{\rm GeV}\leq\sqrt{s} \leq 3.66~{\rm GeV}$ and $2.23~{\rm GeV}\leq\sqrt{s} \leq 3.66~{\rm GeV}$.}
\begin{tabular}{ccccccccc}
\toprule
\multirow{2}{*}{$\sqrt{s}~[GeV]$} &
\multicolumn{2}{c}{BES-III} &
\multicolumn{2}{c}{KEDR} &
\multicolumn{2}{c}{BES} &
\multicolumn{2}{c}{Combined} \\

& pQCD & w/ DVs & pQCD & w/ DVs & pQCD & w/ DVs & pQCD & w/ DVs \\
\midrule
$[1.8; 3.66]$ & --- & --- & $1.0\sigma$ & $0.6\sigma$ & --- & --- & $1.3\sigma$ & $0.6\sigma$ \\
[0.1cm]
$[2.23;3.66]$ & $3.4\sigma$ & $3.2\sigma$ & $1.4\sigma$ & $1.3\sigma$ & $2.2\sigma$ & $2.2\sigma$ & $2.4\sigma$ & $2.0\sigma$ \\
\bottomrule
\end{tabular}
\label{tab:amu_discrepancies} }
\end{table}

\section*{Acknowledgements}

\ni We thank Achim Denig for discussions about the BES-III data and Andrei Kataev for pointing out a typo in a previous version of Eq.~(\ref{eq:deltams}). DB’s work was supported
by the S\~ao Paulo Research Foundation (FAPESP) grant No.~2021/06756-6 and by CNPq
grant No.~303553/2025-1. MC's work was supported by Coordena\c c\~ao de Aperfei\c coamento de Pessoal de N\'ivel Superior (CAPES), Finance Code~001.

\appendix
\section{Renormalon-free gluon-condensate scheme}
\label{app:RFGC}

We give here an overview of the main steps in the construction of the RF GC scheme, which is designed to consistently remove the $u=2$ renormalon ambiguity from the Adler function perturbative series and, as a consequence, from the pQCD expansion of its integrated moments~\cite{Benitez-Rathgeb:2021gvw,Benitez-Rathgeb:2022hfj,Benitez-Rathgeb:2022yqb}. To convey the main ideas, it is sufficient to discuss the implementation in the large-$\beta_0$ limit, avoiding some of the technical aspects that appear in full QCD, that we outline in the end.

The contribution of the $u=2$ IR renormalon to the Borel transformed Adler function in the large-$\beta_0$ limit, associated with condensates of $D=4$, can be written as
\beq
B[\widehat D]_{D=4}(u)= \frac{N_{4}}{2-u} = \frac{32}{3\pi}\frac{e^{10/3}}{2-u},\label{eq:B4}
\eeq
where we considered $\mu^2=Q^2$ and we work in the $\MSbar$ scheme, where $C=-5/3$ in the results of Eq.~(\ref{eq:Borel-Trans-Adler-largebeta}). The constant $N_4=32 e^{10/3}/(3\pi)$ is exactly known in large-$\beta_0$ (but not in full QCD).

This renormalon gives a fixed sign contribution to the Adler function expansion of the form
\beq
\widehat D_{D=4}(\alpha_s) = \frac{32 e^{10/3}}{3\pi} \sum_{n=0}^\infty r_n^{(4)} \alpha_s^{n+1},
\eeq
with coefficients given by
\beq
r_n^{(4)} = \frac{\Gamma(n+1)}{2^{n+1}}\left( \frac{\beta_1}{2\pi} \right)^n.
\eeq
A $D=4$ term in the OPE is associated with the contributions of the $u=2$ pole. In the chiral limit and in large-$\beta_0$, it is given by
\beq
\widehat D^{\rm OPE}_{D=4}(-Q^2) = 2\frac{\langle G^2 \rangle}{Q^4},
\eeq
where the only condensate with dimension $D=4$ is the gluon condensate that we define as $\langle G^2\rangle =\langle \alpha_s G_{\mu\nu}^aG^{\mu\nu,a}\rangle$.

IR poles obstruct the integration contour for the Borel integral and therefore lead to an intrinsic imaginary ambiguity in the true value of the Adler function. This cannot be physical and the GC contribution in the OPE is expected to provide the correction that would eliminate this ambiguity. The RF GC scheme consists in making this explicit by a consistent redefinition of the GC such as to exactly cancel the imaginary ambiguity associated with the $u=2$ renormalon. This redefinition is done in terms of an IR subtraction scale, that we call $R$. This idea is based on previous works that implement similar renormalon subtractions in other contexts~\cite{Pineda:2001zq,Hoang:2009yr,Hoang:2017suc}.

We then redefine the GC, order by order, as
\beq
\langle G^2 \rangle^{(n)} = \langle G^2 \rangle^{\rm RF} -R^4N_g \sum_{n=0}^\infty r_n^{(4)} \alpha_s^{n+1}(R^2) +N_g c_0(R^2),\label{eq:RF-GC}
\eeq
where $N_g=N_4/2$ is the GC norm and the new GC in the RF scheme is $\langle G^2 \rangle^{\rm RF}$. In this redefinition, $\alpha_s(R^2)$ is calculated at the IR subtraction scale but, at the end, one must consistently re-expand it in terms of $\alpha_s(s)$ and powers of $\log\left(R^2/s\right)$. The function $c_0(R^2)$ is introduced such that the new GC is formally RG invariant. A function that accomplishes this goal is, essentially, the Borel sum of the $u=2$ renormalon
\begin{align}
c_0(R^2)
=R^4 \left( \frac{2\pi}{\beta_1} \right) {\rm PV} \int_0^\infty du\, e^{-u/\bar\alpha_s(R^2)} \frac{1}{2-u},
\end{align}
where we introduced the notation $\bar \alpha_s = \beta_1 \alpha_s/(2\pi)$ and `PV' means the Cauchy principal value. This function should {\it not} be expanded, i.e. its role is not to generate terms in an asymptotic expansion, and it must be kept in the closed form given by the Borel integral.

When this redefinition is used in the GC OPE contribution, and one reshuffles all $R$-dependent terms into the perturbative series, the perturbative expansion of the Adler function becomes, with the choice $\mu^2=Q^2=-s$,
\beq
\widehat D^{\rm RF}(s)= \sum_{n=0}^{\infty} r_n \alpha_s^{n+1}(-s) -\frac{R^4}{s^4}N_4 \sum_{n=0}^{\infty}r_n^{(4)} \alpha_s^{n+1}(R^2) +N_4 c_0(R^2).
\eeq
In practice, the sums in the perturbative series have to be truncated at an order $n^*$ (which here is taken to be $n^*=5$). The second term contributes to the cancellation of the $u=2$ contributions from the first sum up to this order. Note that the modification that is done to the perturbative series is minimal, in the sense that differences with respect to the original series are formally $\mathcal{O}(\alpha_s^{n^*+1})$. It is possible to show that the imaginary ambiguity from the pole at $u=2$ is exactly canceled by the imaginary ambiguity of $c_0(R^2)$, and the Adler function is no longer singular at $u=2$. This is a consequence of the fact that the Borel integral is scale independent and to show it, it is crucial to consistently write $\alpha_s(R^2)$ in terms of $\alpha_s(Q^2)$. Finally, another nice feature of the RF GC scheme is that the real part of the Borel sum of the new RF Adler function is the same as before and that the OPE GC correction retains its form
\beq
\widehat D^{\rm OPE, RF}_{D=4}(-Q^2) = 2\frac{\langle G^2 \rangle^{\rm RF}}{Q^4}.
\eeq

Once the RF scheme is implemented, the $u=2$ IR pole is consistently removed, while keeping the Borel integral, i.e. the true value of the Adler function, unaltered. It is now well established that this pole is responsible for more than 99\% of the difference between CIPT and FOPT~\cite{Hoang:2020mkw,Hoang:2021nlz,Golterman:2023oml}. Therefore, in the RF scheme the two series now approach the same value, as shown in Fig.~\ref{fig:delta0-RFPT}.

The implementation of the RF scheme in QCD follows the same steps. The main complications are (i) that the structure of the renormalon singularity is more complicated and (ii) that we do not know the GC norm, $N_g$. To deal with the first difficulty, it is convenient to, as an intermediate step, switch to another scheme for $\alpha_s$, the $C$-scheme~\cite{Boito:2016pwf}, in which the $\beta$ function and the general structure of the renormalon singularities are greatly simplified. One can then, after the the RF scheme is implemented, re-expand $\alpha_s$ to return to the usual $\MSbar$ scheme (as done in Refs.~\cite{Benitez-Rathgeb:2022hfj,Benitez-Rathgeb:2022yqb}). Finally, the value of the GC norm $N_g$ can be estimated with sufficient precision with, e.g., the three methods of Ref.~\cite{Benitez-Rathgeb:2022hfj}. The final uncertainty in $N_g$, which is estimated to be of about 40\%, is not too large and partially cancels in final results due to the two contributions in Eq.~(\ref{eq:RF-GC}) having different signs.

\addcontentsline{toc}{section}{References}
\bibliographystyle{jhep}
\bibliography{References}

\end{document}